\documentclass[12pt]{article}
\usepackage[english]{babel}

\usepackage[letterpaper,top=2cm,bottom=2cm,left=3cm,right=3cm,marginparwidth=1.75cm]{geometry}

\usepackage{amsmath,amsfonts,amssymb}
\usepackage{algpseudocode}
\usepackage{graphicx}
\usepackage[colorlinks=true, allcolors=blue]{hyperref}
\usepackage{subcaption}
\usepackage{authblk}

\newcommand{\RomanNumeralCaps}[1]
\linenumbers

\title{Data-driven spectral turbulence modeling for Rayleigh-B\'enard convection}
\author[1]{Sagy R. Ephrati \footnote{Corresponding author (s.r.ephrati@utwente.nl)}}
\author[1]{Paolo Cifani}
\author[1,2]{Bernard J. Geurts}
\affil[1]{Mathematics of Multiscale Modeling and Simulation, Faculty EEMCS, University of Twente, 7500 AE Enschede,
The Netherlands}
\affil[2]{Multiscale Energy Physics, CCER, Faculty Applied Physics, Eindhoven University of Technology,
5600 MB Eindhoven, The Netherlands}

\newcommand{\sre}[1]{{\color{black} {#1}}}

\begin{document}
\maketitle

\section*{Abstract}
A data-driven turbulence model for coarse-grained numerical simulations of two-dimensional Rayleigh-B\'enard convection is proposed. The model starts from high-fidelity data and is based on adjusting the Fourier coefficients of the numerical solution, with the aim of accurately reproducing the kinetic energy spectra as seen in the high-fidelity reference findings. No assumptions about the underlying PDE or numerical discretization are used in the formulation of the model. We also develop a constraint on the heat flux to guarantee accurate Nusselt number estimates on coarse computational grids and high Rayleigh numbers. Model performance is assessed in coarse numerical simulations at $Ra=10^{10}$. We focus on key features including kinetic energy spectra, wall-normal flow statistics, and global flow statistics. The method of data-driven modeling of flow dynamics is found to reproduce the reference kinetic energy spectra well across all scales and yields good results for flow statistics and average heat transfer, leading to computationally cheap surrogate models. Large-scale forcing extracted from the high-fidelity simulation leads to accurate Nusselt number predictions  across two decades of Rayleigh numbers, centered around the targeted reference at $Ra=10^{10}$.


\section{Introduction}\label{sec:Section1}
Turbulent flows are characterized by the distribution of kinetic energy over a vast range of scales. The nonlinearity in the Navier-Stokes equations ensures that large and small eddies interact with each other, resulting in a wide range of dynamic flow features \cite{pope2000turbulent}. This process transfers kinetic energy from the large energy-containing scales toward smaller scales, until the kinetic energy is finally dissipated by viscosity at the smallest scales. The energy cascade towards the smallest scales yields a significant challenge in computational fluid dynamics in the turbulent regime \cite{geurts2003elements, sagaut2006large}. In order to accurately simulate the flow, the fluid dynamical model should resolve the scales of turbulence down to the Kolmogorov length scale. A direct approach would then require very fine computational grids which is often intractable even with modern-day computing resources. A common way to alleviate the large computational requirements is by reducing the numerical resolution at which an approximate solution to the flow is obtained. To compensate for the lack of refinement of the computational approach, a model term is subsequently added to the governing equations to represent the influence of unresolved dynamics on the resolved components of the flow \cite{geurtsholm2002, piomelli2015, rouhi2016, geurts2022book}.

In this paper, we describe how prior knowledge of flow statistics obtained from an offline fully resolved simulation may be incorporated to construct an online high-fidelity model for coarse numerical simulations. The proposed reduced-order model acts on the numerical solution in spectral space, employing techniques from time series modeling and data assimilation. This model is designed to yield accurate kinetic energy spectra, despite the rather coarse flow representation. We demonstrate the capabilities of this data-driven approach for two-dimensional Rayleigh-B\'enard convection at high Rayleigh number.

Rayleigh-B\'enard convection is a fundamental problem in fluid dynamics, describing buoyancy-driven flows heated from below and cooled from above \cite{kadanoff2001turbulent, kooij2018comparison, ahlers2009heat, kunnen2009}. In particular, thermal convection is meaningful in geophysical processes, such as in describing convective processes in the atmosphere \cite{hartmann2001tropical} or the ocean \cite{marshall1999open}. The large range of scales present in turbulence is also further affected by buoyancy effects. For example, a common phenomenon observed in Rayleigh-B\'enard convection is the formation of spatially coherent structures in large-scale circulation \cite{ahlers2009heat} and, in larger spatial domains, the formation of thermal superstructures \cite{stevens2018turbulent}. On the other hand, a thin boundary layer exists near either of the walls which becomes turbulent and increasingly thinner with growing temperature differences between the walls, i.e., growing Rayleigh number \cite{kraichnan1962turbulent, zhu2018transition}. Properly resolving the boundary layers requires large computational grids and poses a challenge even in two-dimensional Rayleigh-B\'enard convection \cite{zhu2018transition}. This stresses the conundrum of computational fluid dynamics, where one strives to find a balance between simulating flows at modest computational costs without adversely affecting the prediction of flow statistics.

Simulating flows at modest computational costs while retaining a high level of accuracy is the aim of large-eddy simulation (LES) \cite{sagaut2006large, geurts2003elements}. Instead of fully resolving all length scales of the flow, a computationally less intensive approximation is found by coarsening the flow description and simultaneously including a subgrid model to accommodate the loss of explicit finer details in the dynamics. The coarsening is accomplished by spatial filtering, which, through the specification of the filter-width, establishes the required level of refinement that should be included in the numerical simulations. The subgrid model then approximates the effect the unresolved dynamics has on the resolved scales, and serves as a closure for the filtered equations. This approximation depends on both the adopted filter \cite{geurtsholm2003} and selected closure model as well as the choice of discretization and level of coarsening \cite{langford1999optimal, beck2019deep, fedderikbos2007}. 

With the increase of computational resources, direct numerical simulations (DNS) of turbulent flows are achievable to an ever-increasing extent and may serve to generate data from which LES models could be derived. This data-driven LES has been an active topic of research in recent years. For example, the decomposition of unresolved dynamics into fixed global basis functions and corresponding time series yields an efficient approach for which only the latter needs to be modeled.  In \cite{frederiksen2006dynamical} DNS data were employed of the barotropic vorticity equation to model the time series of spherical harmonics as stochastic processes with memory effects, leading to accurate kinetic energy spectra in coarse-grid simulations. Using proper orthogonal decomposition (POD), \cite{ephrati2022computational} showed that applying corrections to coarse-grid numerical simulations may lead to significant error reduction. Machine-learning methods have also been successfully employed to find subgrid models \cite{beck2019deep}, reporting improved results compared to traditional eddy-viscosity models. Examples include using artificial neural networks in two-dimensional decaying turbulence \cite{maulik2019subgrid} and convolutional neural networks in three-dimensional homogeneous isotropic turbulence \cite{kurz2023deep}, yielding improved energy spectra and turbulent fluctuation distributions. \sre{Similar methods have also been applied to two-dimensional Rayleigh B\'enard convection. A reduced-order model is developed in \cite{pandey2022direct} by employing machine learning, where a combined convolutional autoencoder-recurrent neural network \cite{vlachas2020backpropagation} is used to predict the local turbulent heat flux. Accurate probability density functions are obtained for the local heat flux, while a dimensionality reduction to 0.2\% of the original size is achieved for the turbulence data. In a similar vein, \cite{heyder2021echo} study moist Rayleigh-B\'enard convection. The dimensionality of the system is reduced by applying POD to the high-fidelity data, after which the POD coefficients are predicted using echo state networks. Here a good agreement of low-order flow statistics is observed in the reduced-order model.}

Incorporating data into numerical models to improve flow predictions is well-established in geophysical fluid dynamics, where data assimilation has been successfully employed for several decades. The aim is to improve forecasting by minimizing the differences between observed and modeled values while accounting for uncertainties  \cite{ghil1991data, daley1992estimating}. In particular, continuous data assimilation (CDA) aims to nudge the model solution toward an observed reference by means of a feedback control term acting as external forcing \cite{azouani2013feedback, azouani2014continuous}. This concept is also extended to linear stochastic differential equations, arising as the continuous-time limit of the 3DVAR data assimilation algorithm \cite{blomker2013accuracy}, which has been shown to successfully steer coarse-grained numerical solutions of the two-dimensional Navier-Stokes equations towards an observed reference solution. Additionally, the convergence of coarse numerical solutions augmented by CDA to an observed reference solution has been proven \cite{farhat2015continuous} and shown numerically for two-dimensional Rayleigh-B\'enard convection \cite{altaf2015downscaling}.

The purpose of this paper is to combine ideas from data assimilation with large-eddy simulation. In particular, we derive a model term based on statistical quantities from a reference high-resolution simulation and use this as a stand-alone model for coarse numerical simulations. Our proposed method incorporates Ornstein-Uhlenbeck processes in the evolution of the Fourier coefficients of the numerical solution, steering the solution towards an a priori measured statistically steady state. Only three parameters need to be defined for each Fourier mode, outlining the simplicity of the model. The parameters are inferred from data, do not depend on the adopted spatial or temporal discretization, and are defined such that the reference energy spectrum is closely reproduced in the coarse-grid simulations. The resulting prediction-correction scheme is of the form of the diagonal Fourier domain Kalman filter \cite{harlim2008filtering, majda2012filtering} with a fixed prescribed gain. This identification enables future research that combines LES and data assimilation. The same approach has been applied in a recent study of coarse-grid modeling of the two-dimensional Euler equations on the sphere \cite{ephrati2023qeuler}, where a decomposition of the vorticity field into spherical harmonic basis modes was employed in the coarse-grid model.

\sre{The two main results reported in this paper are the following. Firstly, the high-resolution data at Ra=$10^{10}$ can be used successfully to define a forcing acting on large scales of coarse numerical simulations. By improving the energy content in these forced scales in coarsened simulations an accurate estimate of the Nusselt number is obtained. The forcing calibrated at Ra=$10^{10}$ is found to yield accurate Nusselt number estimates across two decades of Rayleigh numbers, from Ra=$10^{9}$ to Ra=$10^{11}$, without the need to re-compute a high-fidelity simulation for each Rayleigh number of interest. Secondly, combining forcing across all resolvable scales with prior knowledge of the heat flux in the domain leads to an `offline/online' approach through which an accurate and effective coarsened model for online real-time predictions can be formulated. The model that yields such accurate coarsening is derived from the high-fidelity offline simulation. The result is a computationally cheap surrogate model for Rayleigh-B\'enard convection, accurately representing the complex behavior up to the smallest resolvable scales on the coarse grid. This is demonstrated here at Ra=$10^{10}$.}

The paper is structured as follows. The governing equations and adopted discretization are described in Section \ref{sec:Section2}. The data-driven model is introduced in Section \ref{sec:Section3}, detailing the forcing in Section \ref{sec:model_description} and complementary heat flux correction in Section \ref{sec:heat_transport_correction}. The model performance for a variety of model configurations is assessed in Section \ref{sec:Section4}. Conclusions are presented in Section \ref{sec:Section5}.

\section{Governing equations and numerical methods}\label{sec:Section2}
In this section, we introduce the governing equations and the simulated case in Section \ref{sec:governing_equations}, the employed numerical discretization in Section \ref{sec:numerical_methods}, and illustrate the effects of severe coarsening, without any modeling correction, on the quality of the solution in Section \ref{sec:coarse_dynamics}.

\subsection{Governing equations} \label{sec:governing_equations}
Rayleigh-B\'enard (RB) convection is described by the incompressible Navier-Stokes equations coupled to buoyancy effects under the Boussinesq approximation. In non-dimensional form, the equations read
\begin{align}
    \frac{\partial \mathbf{u}}{\partial t} + \mathbf{u}\cdot \nabla \mathbf{u} &= \sqrt{\frac{Pr}{Ra}}\nabla^2 \mathbf{u} - \nabla p + T\mathbf{e}_y, \label{eq:momentum}\\
    \nabla\cdot \mathbf{u} &= 0, \label{eq:continuity}\\
    \frac{\partial T}{\partial t} + \mathbf{u}\cdot\nabla T &= \frac{1}{\sqrt{Pr Ra}}\nabla^2 T. \label{eq:energy}
\end{align}
We restrict to two spatial dimensions in this work. Here, $\mathbf{u}$ denotes velocity, $p$ the pressure and $T$ the temperature. We denote by $u$ and $v$ the horizontal and vertical velocity components, respectively. Buoyancy effects are included in the momentum equation by means of the term $T\mathbf{e}_y$, where $\mathbf{e}_y$ denotes the vertical unit vector. The dimensionless parameters that determine the flow are the Rayleigh number $Ra = g\beta\Delta L_y^3/(\nu\kappa)$ and the Prandtl number $Pr=\nu/\kappa$. The Rayleigh number describes the ratio between buoyancy effects and viscous effects and is set to $10^{10}$ in order to set the focus on the challenging high-$Ra$ convection regime. The Prandtl number determines the ratio of characteristic length scales of the velocity and the temperature and is set to 1. The gravitational acceleration is denoted by $g$, the thermal expansion coefficient by $\beta$, the temperature difference between the boundaries of the domain by $\Delta$, the kinematic viscosity by $\nu$, the thermal diffusivity by $\kappa$. The computational domain is rectangular with horizontal length $L_x$ and vertical length $L_y$, which are chosen as 2 and 1, respectively. The domain is periodic for all variables in the horizontal direction and wall-bounded in the vertical direction. No-slip boundary conditions are imposed for the velocity at the walls. The non-dimensional temperature is prescribed at 1 at the bottom wall and 0 at the top wall.

Two-dimensional RB convection is fundamentally different from three-dimensional RB convection.  The main advantage of restricting the flow to two spatial dimensions is that this enables direct numerical simulation (DNS) at a very large Rayleigh number \cite{zhu2018transition}. Additionally, the large-scale circulation that appears in three-dimensional RB convection displays quasi-two-dimensional behavior and shows strong similarities with the large-scale circulation in two-dimensional RB convection \cite{van2013comparison}.

\subsection{Numerical methods}\label{sec:numerical_methods}
The adopted spatial discretization is an energy-conserving finite difference method \cite{vreman2014projection} and is parallelized as in \cite{cifani2018highly}. A staggered grid arrangement is used for the velocity, the pressure is defined at the cell centers, and the temperature is defined on the same grid as the vertical velocity. The latter choice ensures that no interpolation errors occur when computing the buoyancy term in Eq. \eqref{eq:momentum} \cite{van2015pencil}. A uniform grid spacing is used along the horizontal direction whereas a hyperbolic tangent grid profile is adopted along the vertical direction. The non-uniform grid realizes refinement near the walls to resolve the boundary layer. 

Time integration is performed using the fractional-step third-order Runge-Kutta (RK3) scheme for explicit terms combined with the Crank-Nicholson (CN) scheme for implicit terms, as presented in \cite{van2015pencil}. Every time step, from $t^n$ to $t^{n+1}$, consists of three sub-stages, of which the steps are outlined below. The superscript $j$, $j=0, 1, 2,$ denotes the sub-stage, where $j=0$ coincides with the situation at $t^n$. 
In each stage, a provisional velocity $\mathbf{u}^*$ is computed according to  
\begin{equation}
    \frac{\mathbf{u}^* - \mathbf{u}^j}{\Delta t} = \left[ \gamma_j H^j + \rho_j H^{j-1} - \alpha_j\mathcal{G}p^j + \alpha_j\mathcal{A}_y^j \frac{\left(\mathbf{u}^* + \mathbf{u}^j \right)}{2} \right]. \label{eq:provisional_velocity}
\end{equation}
The parameters $\gamma, \rho$, and $\alpha$ are the Runge-Kutta coefficients, given by $\gamma=[8/15, 5/12, 3/4], \rho=[0, -17/60, -5/12]$, and $\alpha=\gamma+\rho$~\cite{cifani2018highly, van2015pencil, rai1991direct}. 
Moreover, $H^j$ is comprised of the discrete convective terms, the discrete horizontal diffusion terms, and the source terms. Here, the only source term is the buoyancy term appearing in the evolution of the vertical velocity. The discrete gradient operator is denoted by $\mathcal{G}$. The discrete vertical diffusion, given by $\mathcal{A}_y$, is treated implicitly. The implicit treatment eliminates the severe viscous stability restriction caused by the use of a nonuniform mesh near the boundary \cite{kim1985application}. Subsequently, the Poisson equation \eqref{eq:poisson} is solved for the pressure to impose the continuity constraint \eqref{eq:continuity},
\begin{equation}
    \nabla^2\phi = \frac{1}{\alpha_j\Delta t}\left(\nabla\cdot\mathbf{u}^*\right). \label{eq:poisson}
\end{equation}
Discretely, this equation takes the form 
\begin{equation}
    \mathcal{L}\phi = \frac{1}{\alpha_j\Delta t}\left(\mathcal{D}\mathbf{u}^*\right) \label{eq:discrete_poisson}.
\end{equation}
Here, $\mathcal{L}$ is the discrete Laplace operator $\nabla\cdot\nabla$ and $\mathcal{D}$ is the discrete divergence operator $\nabla\cdot$. The velocity and pressure are subsequently updated according to
\begin{equation}
    \mathbf{u}^{j+1} = \mathbf{u}^* - \alpha_j\Delta t\left(\mathcal{G}\phi\right) \label{eq:velocity_update}
\end{equation}
and
\begin{equation}
    p^{j+1} = p^j + \phi - \frac{\alpha_j\Delta t}{2 Re}\left(\mathcal{L}\phi \right) \label{eq:pressure_update},
\end{equation}
after which the velocity $\mathbf{u}^{j+1}$ is divergence-free. Time integration of the energy equation \eqref{eq:energy} follows similarly. The newly obtained velocity is used to generate the energy field $T^{j+1}$ analogously to Eq. \eqref{eq:provisional_velocity}.

The convective terms are discretized using the QUICK interpolation scheme \cite{leonard1979stable}. The diffusive terms are discretized using a standard second-order finite difference method, for both spatial directions. Similarly, the discrete gradient $\mathcal{G}$, the discrete divergence $\mathcal{D}$, and the discrete Laplacian $\mathcal{L}$ are defined using finite differences.

\subsection{Altered dynamics under coarsening} \label{sec:coarse_dynamics}
\begin{figure}[h]
    \centering
    \includegraphics[width=0.95\textwidth]{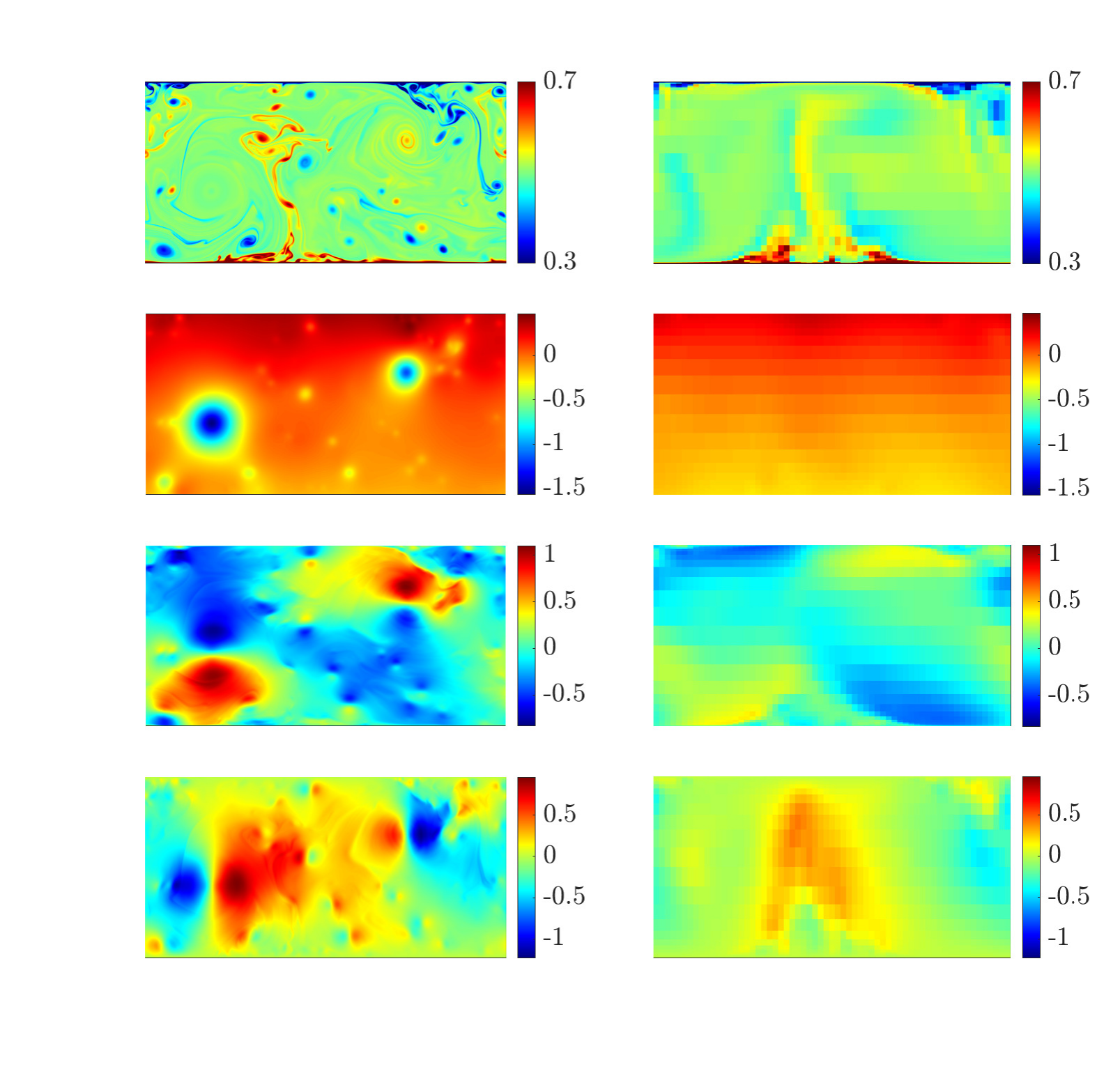}
    \caption{Snapshots of the DNS (4096$\times$2048 grid, left) and coarse no-model simulation (64$\times$32 grid, right) in statistically steady states. From top to bottom, we show temperature, pressure, horizontal velocity, and vertical velocity.}
    \label{fig:dns_no_model_snaps}
\end{figure}

The DNS is carried out on a $4096\times2048$ grid, which has been shown to be a sufficiently high resolution for the chosen Rayleigh number \cite{zhu2018transition}. The coarse-grid numerical simulations will be performed on a $64\times32$ grid. This coarsening introduces significant discretization errors and does not allow for an accurate resolution of the smaller coherent structures present in the flow. The truncation error of the numerical method on this coarse grid introduces artificial dissipation and additional high-pass smoothing native to the discretization method \cite{geurts2005numerically}. Figure \ref{fig:dns_no_model_snaps} shows a snapshot of the DNS and the coarse-grid  simulation, both in statistically steady states, from which the huge implications of such significant coarsening become apparent.

The temperature at the mentioned resolutions is shown in the top row of figure \ref{fig:dns_no_model_snaps}. The coarsened temperature displays only qualitative large-scale agreement with the DNS temperature, both yielding similar plumes of temperature and similar large-scale circulation. Obviously, the persistent small-scale coherent structures visible in the DNS snapshot are lost on the coarse grid. This loss may also be observed for the pressure, depicted in the second row of figure \ref{fig:dns_no_model_snaps}. The high-resolution and low-resolution fields exhibit clear qualitative differences, especially considering the absence of distinct low-pressure regions in the coarse-grid result. A better qualitative agreement between the results at the different resolutions is observed for the velocity components, shown in the bottom rows of figure \ref{fig:dns_no_model_snaps}. At low resolution, qualitatively the same flow patterns can be observed as appear in the high-resolution results, albeit with decreased magnitude. 

The discrepancies between the velocity and temperature fields at the different resolutions clearly pose the challenge we address in this paper. In the next section, we therefore specify our new forcing approach which aims to rectify the observed differences largely. The extent to which this new approach is successful will be scrutinized in Section \ref{sec:Section4}.

\section{Spectrum-preserving forcing}\label{sec:Section3}
In this section, we describe a data-driven forcing to augment coarsened numerical simulations of statistically steady states. The high-resolution and low-resolution snapshots presented in the previous section hinted at the need of introducing explicit forcing to more accurately represent average flow patterns on coarse computational grids.
Here, we propose a model to reproduce the kinetic energy spectra in coarse numerical simulations.

The model parameters are extracted from the \sre{reference data, obtained from} a sequence of 8040 solution snapshots each separated by $0.05$ time units of a DNS performed at a 4096$\times$2048 computational grid. With these numerical data we achieve sufficiently many snapshots to reliably recover statistical properties of the flow, which is a pre-requisite for our model development. The Fourier components of horizontal cross-sections of the solution are computed for each snapshot, yielding time series for each streamwise wavenumber at each $y$-coordinate. 
The magnitudes of these complex time series yield mean values, variances, and correlation times that are used as model parameters. We next present the main steps in the construction of this model.

\subsection{Reconstructing the energy spectra} \label{sec:model_description}
The momentum equation \eqref{eq:momentum} \sre{and temperature equation \eqref{eq:energy}} can be written as a system of complex ODEs for the mode coefficients through projection onto a Fourier basis. \sre{In what follows, we only describe a spectrum-preserving forcing for the momentum. The same derivation is used to define a forcing for the temperature.} We note that a spectral decomposition of the velocity \sre{and temperature} fields is not straightforward due to the presence of boundaries along one spatial dimension. 
Therefore, we restrict ourselves to one-dimensional periodic horizontal profiles to ensure that the Fourier series is well-defined. This choice enables the model to explicitly identify wall-induced features of the flow. Alternatively, after taking into account the nonuniform grid spacing in the wall-normal direction the velocity field can be decomposed by applying a sine transform in the wall-normal direction or by periodically extending the domain and applying a Fourier transform, but these approaches are not pursued in this study. 

For a fixed vertical coordinate $y_l$, the profile $\mathbf{u}(x, y_l, t)$ is decomposed into Fourier modes and corresponding complex coefficients $c_{k, l}$, where $k$ denotes the wavenumber. The coefficients satisfy the system of ODEs \begin{equation}
    \frac{\text{d}c_{k, l}}{\text{d}t} = L\left(\mathbf{c}, k, l\right), \hspace{3mm} k=0,\ldots,N_x/2, \label{eq:spectral_coeff}
\end{equation}
where $L$ involves the spectral representation of of $\mathbf{u}\cdot\nabla$, $\nabla$, $\nabla^2$ and the source term in Eq. \eqref{eq:momentum}. We have already assumed a finite truncation of the number of Fourier modes in this formulation. The vector $\mathbf{c}$ is comprised of all Fourier coefficients up to the largest resolvable wavenumber. 

To arrive at a spectrum-preserving forcing, it is sufficient to consider only the magnitude of the Fourier coefficients $|c_{k, l}|$. These evolve according to a system of ODEs \begin{equation} 
\frac{\text{d}|c_{k, l}|}{\text{d}t} = L_r(\mathbf{c}, k, l), \hspace{3mm} k=0, \ldots, N_x/2, \label{eq:spectral_abs}
\end{equation}
where $L_r$ is introduced to simplify notation. In the model implementation, the definition and explicit computation of $L$ and $L_r$ is not required. Regarding $|c_{k, l}|$ as a stationary stochastic process, we observe that $\mathbb{E}\left(|c_{k, l}|^2\right)$ describes the mean energy content of the $k^\mathrm{th}$ Fourier mode at height $y_l$. By the definition of variance, we find that \begin{equation}
    \mathbb{E}\left(|c_{k, l}|^2\right) = \mathrm{var}(|c_{k, l}|) + \mathbb{E}\left(|c_{k, l}|\right)^2. \label{eq:variance}
\end{equation}
Thus, it is sufficient to obtain an accurate mean value and variance of $|c_{k, l}|$ to achieve the desired energy content in the $k^\mathrm{th}$ Fourier mode. We aim to reproduce the mean value and variance of $|c_{k, l}|$ by augmenting Eq. \eqref{eq:spectral_abs} with an Ornstein-Uhlenbeck (OU) process,
\begin{equation}
    \text{d}|c_{k, l}| = L_r(\mathbf{c}, k, l)\text{d}t + \frac{1}{\tau_{k, l}}\left(\mu_{k, l} - |c_{k, l}|\right)\text{d}t + \sigma_{k, l}\text{d}W_{k,l}^t \label{eq:spectral_model},
\end{equation}
where $\mu_{k, l}$, $\tau_{k, l}$ and $\sigma_{k, l}$ are statistical parameters inferred from a sequence of snapshots of the reference solution. The desired mean value of $|c_{k, l}|$ is given by $\mu_{k, l}$, the term $\sigma_{k,l}\text{d}W_{k,l}^t$ serves to match the measured variance. Here, $\text{d}W_{k,l}^t$ is a general stochastic process which can be tailored to the data \cite{ephrati2023data}, although the common choice is to let it be a Gaussian process with a variance depending on the time step size \cite{higham2001algorithmic}. We adopt the latter and include the variance scaling in the definition of $\sigma_{k,l}$. The forcing strength is determined by a timescale $\tau_{k, l}$ and can be specified per $k, l$ separately. Detailed specifications of the adopted values of $\mu_{k,l}, \sigma_{k,l},$ and $\tau_{k,l}$ follow shortly. The combination of the original dynamics and the feedback control, including the stochastic term, arises as the continuous-time limit of the 3DVAR data assimilation algorithm \cite{blomker2013accuracy}. The model assumes that the unresolved dynamics can be accurately represented by independent stochastic processes. Interactions in the vertical direction are included via the fully resolved simulations which are basis to the forcing.
We will demonstrate that this model is capable of producing accurate results. 

The model will be applied as a prediction-correction scheme. In the case of the RK3 scheme adopted in this work, the following steps are performed for each RK sub-stage. First, the provisional velocity $\mathbf{u}^*$ is computed as in Eq. \eqref{eq:provisional_velocity}. The Fourier coefficients $\tilde{c}_{k, l}$ are computed for this provisional velocity, after which the model is applied in the form of a correction. \sre{In terms of the Fourier coefficients, the algorithm reads} \begin{align}
	\tilde{c}_{k, l}^{j+1} &= c_{k, l}^j + \Delta t \alpha_j L(\mathbf{c}^j, k, l) &\text{(time integration)},\label{eq:prediction}\\
    |c_{k, l}^{j+1}| &= |\tilde{c}_{k, l}^{j+1}| + \frac{\alpha_j \Delta t}{\tau_{k, l}}\left(\mu_{k, l} - |\tilde{c}^{j+1}_{k, l}|\right) + \sigma_{k, l} \Delta W_{k, l} &\text{(correction)}. \label{eq:nudge}
\end{align}
\sre{Here, Eq. \eqref{eq:prediction} is a simplified notation of the time marching method presented in Section \ref{sec:numerical_methods} and the superscript $j, j=0,1,2,$ denotes the RK sub-stage.} The values of $\Delta W_{k, l}$ are samples from a standard normal distribution, independently drawn for each $k$ and $l$. These are determined for each time step and kept constant throughout the sub-stages comprising the time step. The correction only affects the magnitudes of the Fourier coefficients, the phases of $c^{j+1}_{k, l}$ are the same as those of $\tilde{c}^{j+1}_{k, l}$. Velocity fields are subsequently obtained by applying the inverse Fourier transform to the corrected coefficients $c^{j+1}_{k, l}$. After this procedure, the Poisson equation \eqref{eq:discrete_poisson} is solved using the newly obtained velocity fields and the remaining steps of the sub-stage are completed. Applying the model before solving the Poisson equation ensures that the flow is incompressible at the end of each RK sub-stage. \sre{The entire algorithm, with the exception of solving the Poisson equation, may also be applied to the temperature equation.} This prediction-correction algorithm has the additional benefit that the model can be easily implemented into already existing computational methods.  \sre{Applying the correction once every few time steps instead of every time step is common in data assimilation, where real-time data becomes available sequentially and may not be available at each time step of the numerical simulation. In the current study, all data is collected before performing a coarse numerical simulation and serves to define a forcing term aiming to counteract coarsening effects. The subsequent application of the correction does not noticeably add to the computational effort and is hence applied at each RK sub-stage.}

The correction \eqref{eq:nudge} will be referred to as \textit{nudging}. We distinguish between stochastic nudging, which is described by Eq. \eqref{eq:nudge}, and deterministic nudging, for which the stochastic term in Eq. \eqref{eq:nudge} is omitted. We define a mean $\mu_{k,l, \mathrm{stoch}}$ and $\mu_{k,l, \mathrm{det}}$ for these methods, respectively, and specify these below.

The mean $\mu_{k,l}$ is specified such that the desired energy content is reproduced for small values of $\tau_{k,l}$. The magnitude of the coefficients is fully determined by the model in the limit of small $\tau_{k,l}$ and, as a result, Eq. \eqref{eq:variance} can be used to derive the mean $\mu_{k,l}$ \sre{in this limit}. To attain the desired energy contents when using stochastic nudging, we require that $\mu_{k,l, \mathrm{stoch}}=\mathbb{E}\left(|c_{k,l}|\right)$. In the case of deterministic nudging with small $\tau_{k,l}$, the magnitudes of the coefficients remain constant at the value of $\mu_{k,l, \mathrm{det}}$. Thus, the variance of $|c_{k, l}|$ is set to zero and we require that $\mu_{k,l,\mathrm{det}}=\sqrt{\mathbb{E}\left(|c_{k, l}|^2\right)}$. \sre{We note that the limit of small $\tau_{k,l}$ is used to derive the model parameters, but the actual measured values of $\tau_{k,l}$ from the high-resolution data will be (considerably) larger. As a result, the magnitudes obtained using the model will be a combination of the coarse discretization and the high-resolution data.}

\sre{Treating the evolution operator $L_r$ of the magnitudes of the Fourier coefficients in Eq. \eqref{eq:spectral_abs} as the identity operator allows us to specify the noise magnitude $\sigma_{k,l}$. This assumption is used in the 3DVAR data assimilation algorithm \cite{lorenc2005does} and is valid in statistically stationary states and on short assimilation intervals. This allows replacing $|\tilde{c}^{j+1}_{k,l}|$ by $|c^j_{k,l}|$ and reduces Eq. \eqref{eq:nudge} to a first-order autoregressive (AR$(1)$) process.} We observe that the drift coefficient is $(1-\alpha_j\Delta t/\tau_{k,l})$ and assume that the sample variance $s^2_{k, l}$ is known from the high-fidelity data for every $k, l$. The noise magnitude follows by matching the variance of the AR(1) process with the sample variance, leading to the expression \begin{equation}
    \sigma_{k,l} = s_{k,l}\sqrt{1 - \left(1 - \frac{\alpha_j\Delta t}{\tau_{k,l}}\right)^2}.
\end{equation}
\sre{In this paper, the time scale $\tau_{k,l}$ will be defined as the correlation time of the corresponding Fourier coefficient, measured from the high-resolution data.} We note that, with the adopted definitions of $\mu_{k,l}$ and $\sigma_{k,l}$, the time scale $\tau_{k, l}$ can take on a range of values whilst still yielding accurate energy spectra. Robustness of the model under variations of $\tau_{k, l}$ is studied in Section \ref{sec:tau_scaling}. In fact, $\tau_{k, l}$ can take on any positive value larger than or equal to $\alpha_j \Delta t$. Small lengthscales are expected to yield a small value of $\tau_{k,l}$, resulting in an increased weight towards the model term and an increased noise magnitude for the stochastic term in the nudging. The model term will have a decreased weight at scales for which a large $\tau_{k,l}$ is measured, which is often observed for large spatial scales. These would correspondingly follow the deterministic resolved dynamics more closely.

The proposed prediction-correction method is of the same form as Fourier domain Kalman filtering \cite{majda2012filtering, harlim2008filtering}. \sre{By defining a prediction and an observation, the approach can be understood as a steady-state filter with a prescribed gain factor and can be placed in the context of data assimilation. The only necessary parameters are the means $\mu_{k,l}$, the variances $\sigma^2_{k,l}$ and the correlation times $\tau_{k,l}$, which are interpreted as follows. } At each sub-stage of the RK3 scheme, the prediction is obtained by evolving the velocity  and temperature fields according to the coarse-grid discretization. The `observation' then consists of velocity or temperature fields sampled from the reference statistically stationary state. For stochastic nudging, these are velocity or temperature fields where the magnitudes of the Fourier coefficients are drawn from normal distributions with mean $\mu_{k,l,\mathrm{stoch}}$ and variance $\sigma^2_{k,l}$. In the deterministic case, the observation consists of these fields with Fourier coefficients of prescribed magnitudes $\mu_{k,l, \mathrm{det}}$. \sre{The prediction is subsequently nudged towards the observation by correcting the predicted magnitudes of the Fourier coefficients. The weight of the model, often referred to as the `gain', is determined by $\alpha_j\Delta t / \tau_{k, l}$, for each $k, l$ separately.}

\subsection{Heat transport correction} \label{sec:heat_transport_correction}
The heat transport in the turbulent flow is described by the Nusselt number and is considered the key response of the system to the imposed Rayleigh number \cite{ahlers2009heat}. The definition of the Nusselt number that we adopt here is \begin{equation}
    Nu = 1 + \sqrt{Pr Ra}\langle v T\rangle_\Omega, \label{eq:Nu_vol}
\end{equation}
which is well-suited for use on coarse computational grids. An alternative definition of $Nu$ involves a gradient of temperature, which is more sensitive to coarse-graining. In Eq. (\ref{eq:Nu_vol}) $\Omega$ denotes the domain with area $|\Omega|$ and $\langle\cdot\rangle_\Omega$ denotes the domain average. It is clear from definition \eqref{eq:Nu_vol} that $vT$ needs to be modeled accurately to recover skillful predictions of the heat flux. To achieve this, we propose a constraint to be used in conjunction with the model described in Section~\ref{sec:model_description}.

The volume average in \eqref{eq:Nu_vol} is comprised of averages of the heat flux along horizontal cross-sections of the domain. For a fixed vertical coordinate $y_l$, we denote the heat flux along this cross-section by $\langle v T \rangle_l$. Along this cross-section, we indicate the Fourier coefficients of the velocity and temperature with a hat symbol $\hat{\cdot}$ and observe that \begin{equation}
    (\widehat{vT})_0 = \sum_{k}\widehat{v}^*_k\widehat{T}_k,
\end{equation}
where the subscript $k$ signifies the $k^\mathrm{th}$ Fourier coefficient.
The subscript $0$ indicates that we consider the zeroth mode of the Fourier series, which by definition equals the value of $\langle v T\rangle_l$. We assume that a mean heat flux along the horizontal cross-section is known from the reference high-resolution data and denote this value by $F_l$. Subsequently, the heat flux along the horizontal cross-section in a coarse numerical simulation is corrected by minimizing the error $\|F_l - \sum_k \widehat{v}_k^* \widehat{T}_k \|^2$ with respect to $T$. Here, we minimize the error by varying the phases of the Fourier coefficients $\widehat{T}_k$. We alter the temperature instead of the vertical velocity so that the velocity field remains divergence-free. Adapting the phases only ensures that the spectrum of the temperature along the horizontal cross-section is invariant under the heat transport correction. In total, the heat flux correction is an extension of the nudging procedure \eqref{eq:nudge}. It enables a correction of the temperature field solely based on a statistic of the reference solution, rather than on a dynamic equation. In doing so, the dependence between the vertical velocity and the temperature is taken into account in the nudging procedure. Thus, applying the heat flux correction ensures an improved average Nusselt number estimate and is therefore expected to improve the accuracy of the numerical solutions.

The error $\|F_l - \sum_k \hat{v}_k^* \hat{T}_k \|^2$ is minimized using a gradient descent algorithm. We note that the correction may in principle yield an arbitrarily good approximation of the reference heat flux, but this is not guaranteed to produce physically relevant results. Instead of aiming for an exact agreement of the mean heat flux, we apply
the gradient descent algorithm until the heat flux in the horizontal cross-section is within a 10\% margin of the reference value. This serves to demonstrate the added value of the correction. Preliminary tests have shown that this already improves the heat flux significantly without qualitatively altering the temperature field. In the next section, coarse numerical simulations are performed both with and without the heat flux correction. The optimization of this procedure is beyond the scope of this paper.

\section{Model performance}\label{sec:Section4}

In this section, we apply the model in eight different configurations to numerical simulations on the coarse grid. The configurations are listed in table \ref{tab:config} and differ in the variable that is being forced, the wavelengths at which the forcing is applied, and whether the forcing is deterministic or stochastic. These configurations will be referred to as M0-7, inspired by the nomenclature used in the comparison of LES models in \cite{vreman1997large}. Here, the wavenumbers at which the forcing is applied are chosen as $l\leq 5$ and $l\leq 32$. The former implies that the model only explicitly acts on the large scales of motion and the latter implies that all resolved scales are directly affected by the model. This set of configurations is chosen to distinguish between the effects of large-scale forcing and small-scale forcing, deterministic forcing and stochastic forcing, and the choice of the forced variable. The model simulations are run with a time step size $\Delta t = 0.02$. \sre{Recall that the filtered DNS provides the reference solution.}

\begin{table}
    \centering
    \begin{tabular}{l l l l l}
    & Model & Forced variable & Wavenumbers & Curve \\
    \hline
    & Filtered DNS & & & solid \\
    M0 & No model &  & & dashed \\
    M1 & deterministic & $u$ & $k\leq 5$ & dash-dotted \\
    M2 & deterministic & $u$ & $k\leq 32$ & dotted \\
    M3 & stochastic & $u$ & $k\leq 5$ & $*$ \\
    M4 & stochastic & $u$ & $k\leq 32$ & $+$ \\
    M5 & deterministic & $T$ & $k\leq 32$ & $\times$ \\
    M6 & deterministic & $u, T$ & $k\leq 32$ & $\square$ \\
    M7 & deterministic, heat flux correction & $u, T$ & $k\leq 32$ & $\lozenge$ \\
    \end{tabular}
     \caption{Model configurations used in the coarse numerical simulations.}
    \label{tab:config}
\end{table}

\begin{figure}
\includegraphics[width=\textwidth]{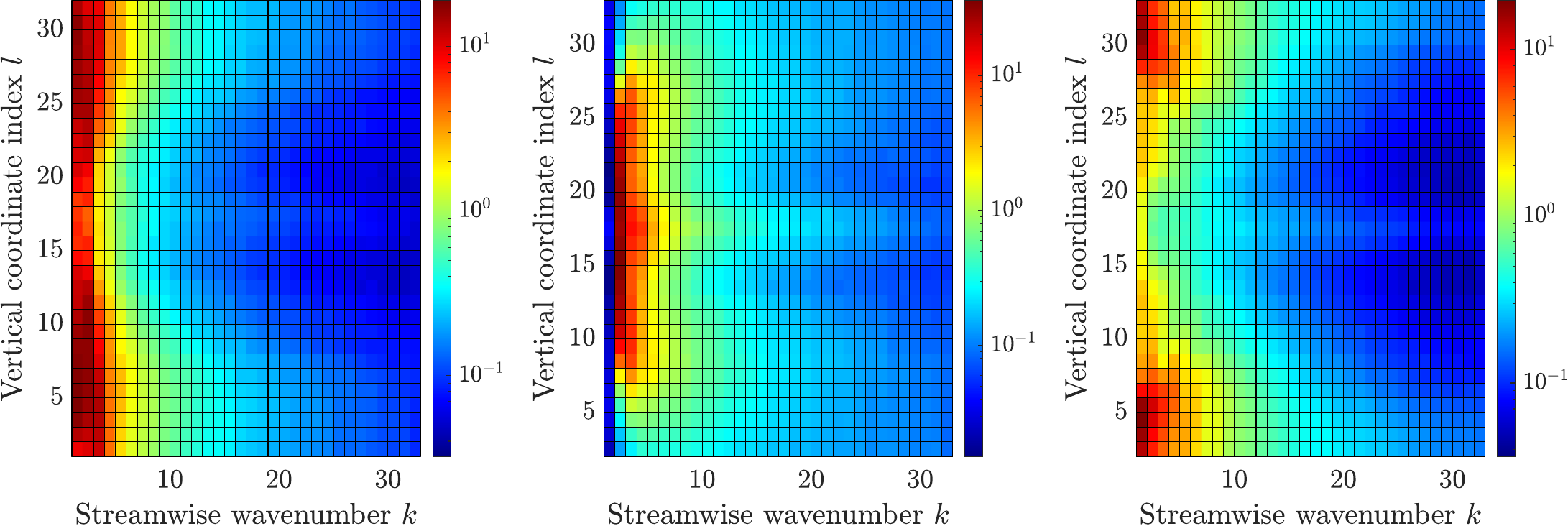}
\caption{Correlation times $\tau_{k,l}$ of the Fourier coefficients for the horizontal velocity (left), vertical velocity (middle), and temperature (right) measured from the high-fidelity data.}
\label{fig:tau_plots}
\end{figure}

\sre{Fig. \ref{fig:tau_plots} shows the correlation times of the Fourier coefficients $\tau_{k,l}$ for each pair $k, l$ for the horizontal velocity, the vertical velocity, and the temperature, measured from the high-fidelity data. As expected, large streamwise wavenumbers, i.e., structures with small length-scales, correspond to small correlation times. Conversely, small wavenumbers, i.e., large length-scale structures, correspond to large correlation times. This confirms that the model contribution to the dynamics of the Fourier coefficients at small scales is stronger than at large scales. The horizontal velocity and the temperature generally yield smaller correlation times in the bulk of the domain compared to the region near the walls, whereas the opposite is observed for the vertical velocity. There is a notable difference at the zero-wavenumber correlation time for the vertical velocity, which is due to the fact that the average vertical velocity is always zero. Finally, we note that $\tau_{k,l}$ is observed to be nowhere below $\alpha_j\Delta t$, thus implying that the magnitudes of the Fourier coefficients do not depend solely on the model in any case, but also take contributions from the discretized equations directly. }

We first provide in Section \ref{sec:qualitative_comparison} an impression of the qualitative improvements obtained when applying the model. In the ensuing subsections, a detailed quantitative comparison is carried out. Several quantities will be compared with the filtered DNS data to gain insight into the quality of the model. In Section \ref{sec:spectra}, we first verify that the model approximates the average energy spectra of the filtered DNS by comparing the spectra of the velocity and the temperature near the wall and in the core of the domain. In Section \ref{sec:flow_statistics}, the mean temperature, the mean heat flux, and the root-mean-square deviation (rms) are measured as a function of wall-normal distance and compared to the reference. Finally, global flow statistics such as the total kinetic energy and the Nusselt number are examined in Section \ref{sec:global_stats}.

The rms, mean temperature and mean heat flux rely on averages along horizontal cross-sections of the domain. For a fixed value of $y$, we adopt the following definition \begin{align}
    \mathrm{rms}(f, y, t) = \left[\frac{1}{|A|}\int_A \left(f(x, y, t) - \langle f(x, y, t)\rangle_A \right)^2\text{d}A \right]^{1/2},
\end{align}
where $\langle\cdot\rangle_A$ denotes the average over the horizontal cross-section with length $|A|$ and $f$ is the field of interest.  The mean temperature and heat flux are computed as the mean $\langle\cdot\rangle_A$ of the corresponding fields. The global kinetic energy will be computed as \begin{equation}
    \mathrm{KE} = \int_\Omega \frac{1}{2}\left(u^2 + v^2  \right)\,\text{d}\Omega \label{eq:KE_def}
\end{equation}
and the Nusselt number follows from definition \eqref{eq:Nu_vol}. 

Our interest lies in the time average of the aforementioned quantities.
The quality of coarse-grid models is therefore measured by comparing averaged quantities rather than instantaneous quantities \cite{langford1999optimal, vreman1997large}. The energy spectra, rms values and mean temperature, and heat flux will be measured after the coarse-grid numerical simulations have reached a statistically steady state. The global quantities of interest are illustrated using a rolling average over time.

\subsection{Qualitative model performance} \label{sec:qualitative_comparison}
\begin{figure}
\includegraphics[width=\textwidth]{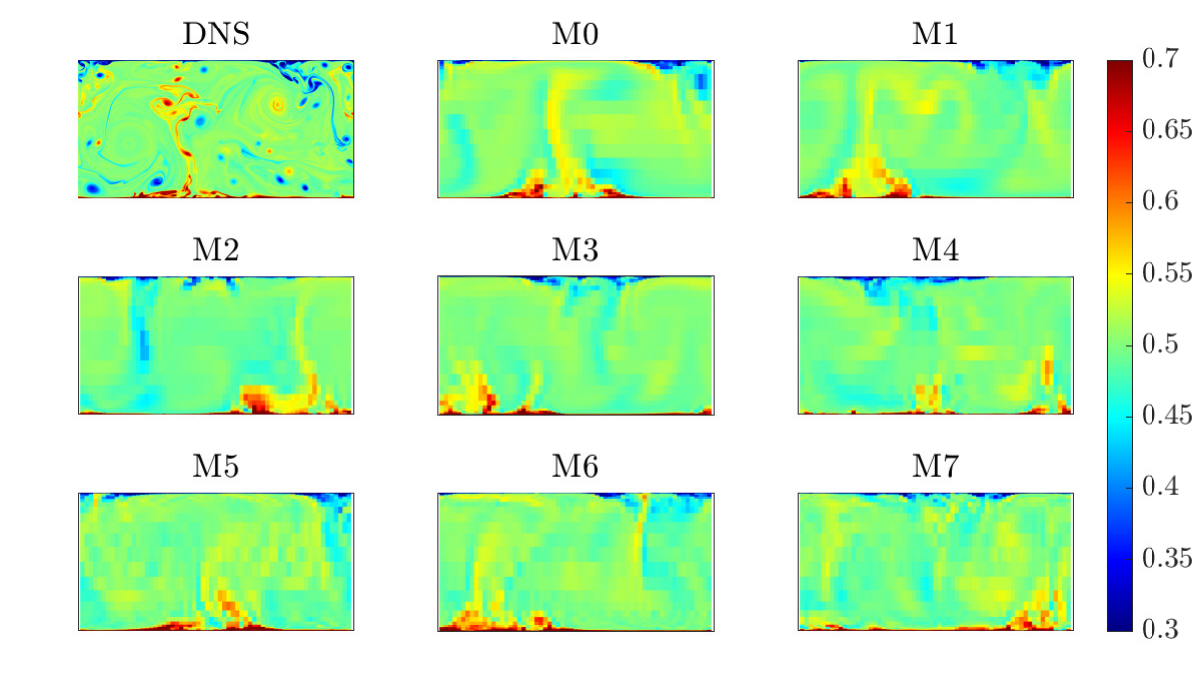}
\caption{Temperature fields in statistically steady states. Shown are the reference solution and the results obtained with coarse numerical simulations M0-7. The color scheme is the same as used for the temperature fields shown in figure \ref{fig:dns_no_model_snaps}.}
\label{fig:snapshot_array_temperature}
\end{figure}

\begin{figure}
\includegraphics[width=\textwidth]{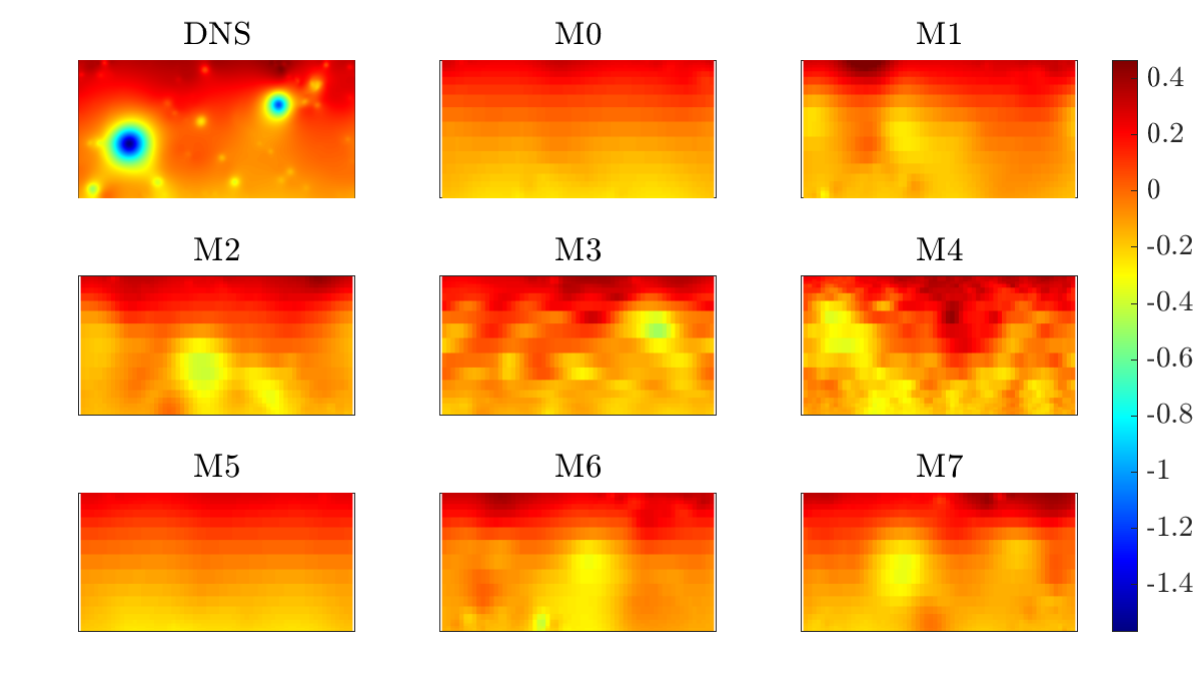}
\caption{Pressure fields in statistically steady states. Shown are the reference solution and the results obtained with coarse numerical simulations M0-7. The color scheme is the same as used for the pressure fields shown in figure \ref{fig:dns_no_model_snaps}.}
\label{fig:snapshot_array_pressure}
\end{figure}

\begin{figure}
\includegraphics[width=\textwidth]{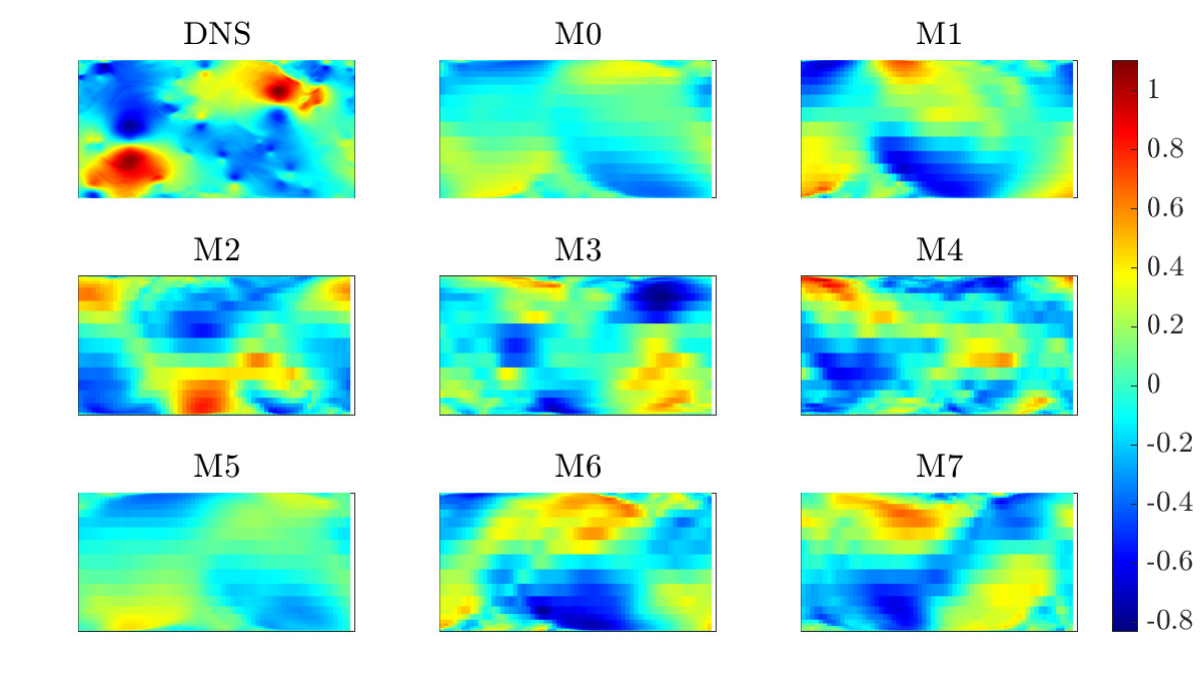}
\caption{Horizontal velocity fields in statistically steady states. Shown are the reference solution and the results obtained with coarse numerical simulations M0-7. The color scheme is the same as used for the horizontal velocity fields shown in figure \ref{fig:dns_no_model_snaps}.}
\label{fig:snapshot_array_horizontalvel}
\end{figure}

\begin{figure}
\includegraphics[width=\textwidth]{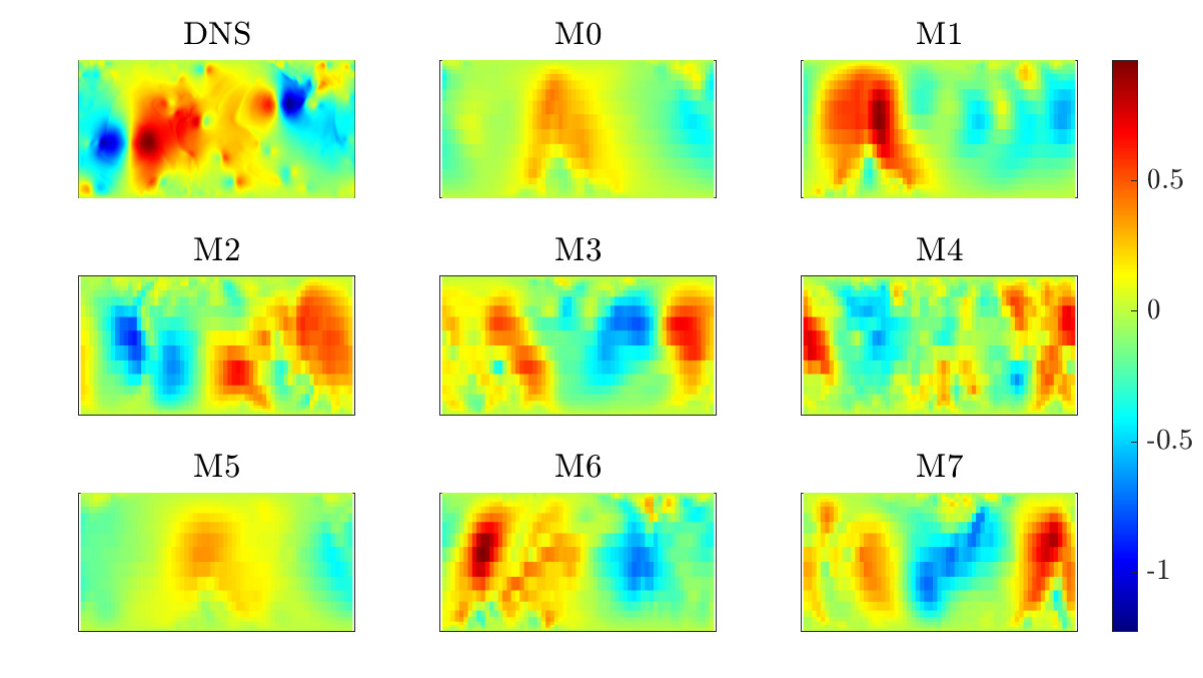}
\caption{Vertical velocity fields in statistically steady states. Shown are the reference solution and the results obtained with coarse numerical simulations M0-7. The color scheme is the same as used for the vertical velocity fields fields shown in figure \ref{fig:dns_no_model_snaps}.}
\label{fig:snapshot_array_verticalvel}
\end{figure}

A qualitative comparison of the model configurations M0-7 is given in figures \ref{fig:snapshot_array_temperature} to \ref{fig:snapshot_array_verticalvel} \sre{by means of instantaneous snapshots in of the obtained statistically stationary states}. In these figures, the snapshots of the DNS and of M0 are the same as depicted in figure \ref{fig:dns_no_model_snaps}. A comparison of the temperature fields in statistically steady states is provided in figure \ref{fig:snapshot_array_temperature}. Here, we observe that the configurations M1-4 do not lead to significant qualitative changes in the temperature field when compared to the no-model configuration M0. In these configurations, the temperature is not explicitly forced and suffers from artificial dissipation inherent to the coarsening. The model configurations M5-7, in which the temperature is forced directly, display more pronounced small-scale features. At the same time, the large-scale circulation pattern is still visible in these results. In addition, from the results of M6 and M7 we conclude that applying the heat flux correction does not lead to qualitatively different temperature fields.

The pressure fields of the corresponding solutions are shown in figure \ref{fig:snapshot_array_pressure}. Here, we recall that any detail obsereved in the DNS pressure field is lost in the coarse no-model result M0. No improvements are observed in the pressure field when only the temperature is explicitly forced, as is done in M5. The remaining model configurations all yield a distinct qualitative improvement in the pressure fields. In particular, only applying a large-scale velocity correction already qualitatively changes the pressure field. This is observed for the deterministic and the stochastic forcing, given by M1 and M3, respectively. The addition of forcing the velocity at small scales or simultaneously forcing the temperature does not yield additional significant changes. A noticeable difference exists between the deterministic and stochastic methods. As becomes clear from M3-4, the random forcing leads to a fragmentation of the coherent structures in the pressure field. 

The horizontal velocity fields and vertical velocity fields are provided in figures \ref{fig:snapshot_array_horizontalvel} and \ref{fig:snapshot_array_verticalvel}, respectively. We observe that all coarse numerical solutions display agreement with the DNS in terms of large-scale coherent structures. Nonetheless, artificial dissipation leads to an underestimate of the velocity magnitude in cases M0 and M5. This suggests that only forcing the temperature is not sufficient for accurately reproducing the velocity fields. The other cases indicate that explicitly forcing the velocity leads to accurate velocity magnitudes.

\subsection{Energy spectra} \label{sec:spectra}
\begin{figure}
    \centering
    \includegraphics[width=\textwidth]{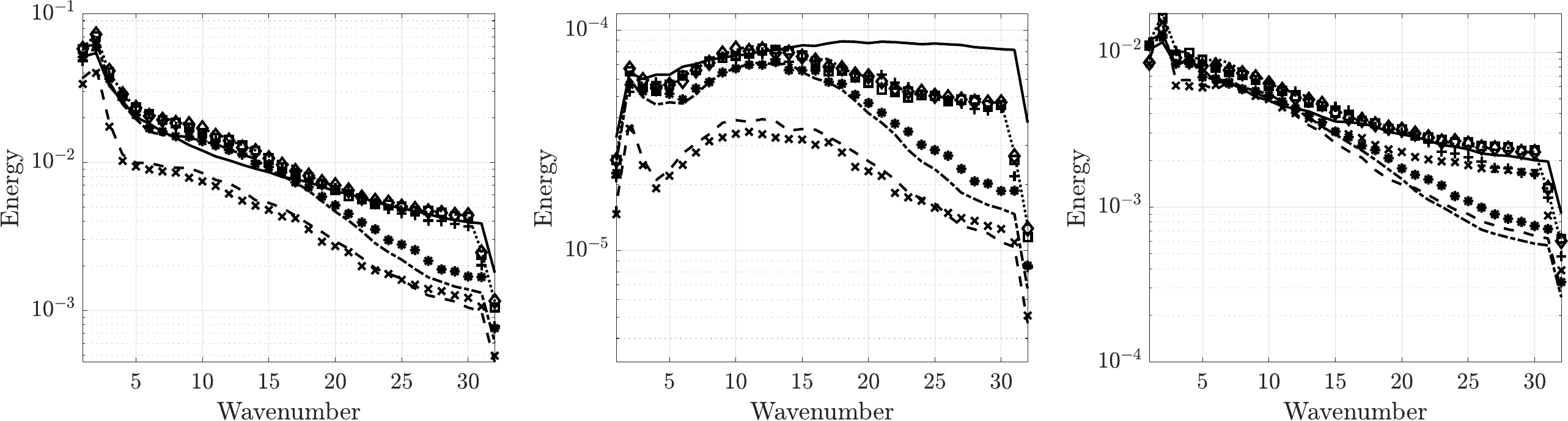}
    \includegraphics[width=\textwidth]{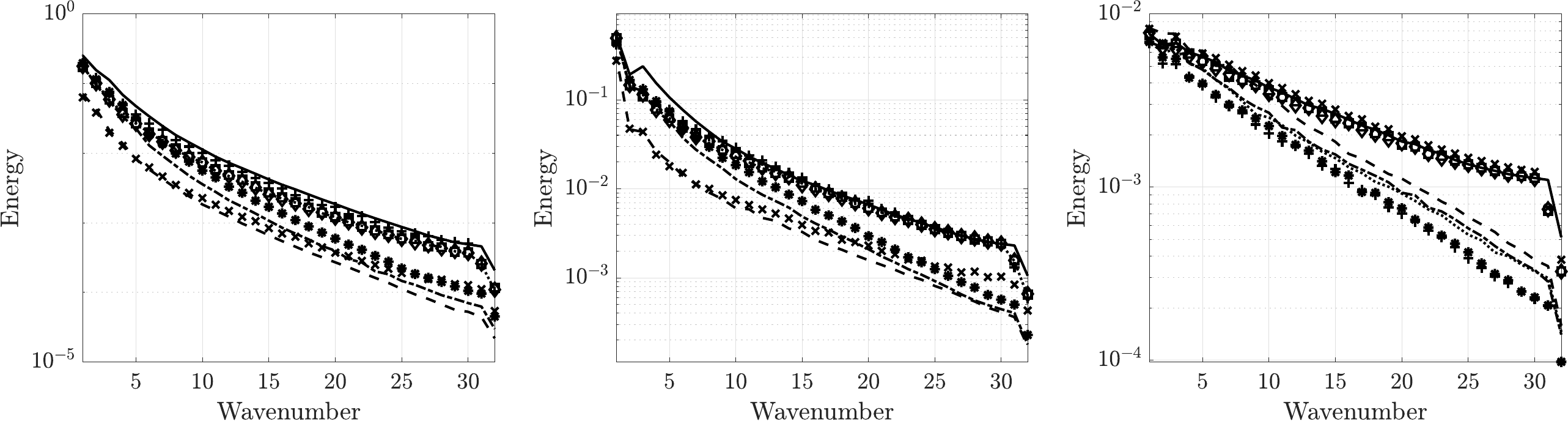}
    \caption{Time-averaged energy spectra measured along horizontal cross-sections of the domain for the horizontal velocity (left column), vertical velocity (middle column), and temperature (right column). The cross-sections are taken near the bottom wall (top row) and the core of the domain (bottom row). The cross-sections are taken at $y=8.5\times 10^{-4}, y=5.5\times10^{-1}$ for the horizontal velocity and at $y=5.0\times 10^{-4}, y=5.0\times10^{-1}$ for the vertical velocity and the temperature. The solid lines show the average spectra of the filtered DNS, the model results are displayed using the symbols in table \ref{tab:config}.}
    \label{fig:energy_spectra}
\end{figure}

We now establish that the model proposed in Section \ref{sec:model_description} improves the average energy spectra of the forced variables. The average energy spectra of the velocity components and the temperature are shown in figure \ref{fig:energy_spectra}, displaying the spectra along a horizontal cross-section near the wall and in the center of the domain. Both near the wall and in the center of the domain, respectively shown in the top and bottom row, the no-model M0 results exhibit significant differences compared to the filtered DNS. The measured energy levels of the velocity are too low with M0 at all resolved scales. In contrast, a significant discrepancy in the temperature spectra is observed only for wavenumbers larger than 10.

The discrepancies in the spectra of M0 and the reference are attributed to artificial dissipation caused by the coarsening. In particular, the numerical dissipation affects both the velocity and the temperature spectra at higher wavenumbers. Through the nonlinear interactions in the momentum equation the velocity is adversely affected at all wavenumbers. This is further corroborated by the results of M2 and M4, where all available lengthscales are forced only for the velocity. In the core of the domain, where the coarsening is strongest, these results display accurate velocity spectra but yield no improvement in the temperature spectra, suggesting that the temperature still suffers from artificial viscosity in these cases. Apparently, the improvements in the velocity spectra influence the prediction of the temperature only to a small degree.

The large-scale velocity forcing applied in M1 and M3 yields improved velocity energy levels at low wavenumbers. However, the improvement gradually vanishes at higher wavenumbers. These configurations exhibit no improvement in the temperature spectra. The cases M2 and M4 lead to an improved agreement on the velocity spectra at all wavenumbers in the center of the domain, establishing the spectrum-reconstructing property of the model described in Section \ref{sec:model_description}. Nonetheless, all models underestimate the large-scale energy in the center of the domain. At these scales, the measured correlation time $\tau$ is large and therefore the model contribution is limited.

Near the wall, the horizontal velocity is accurately represented at all wavenumbers despite the fact that the energy of the vertical velocity deviates from the reference for wavenumbers larger than 15. The temperature spectra for M2 and M4 near the wall show good agreement with the reference. Comparing this to the results of M1 and M3 indicates that the prediction of near-wall temperature is improved by the forcing of small-scale velocity despite no explicit forcing being applied to the temperature. No improvement is observed for these cases in the center of the domain, which we attribute to artificial dissipation.

The velocity spectra show no significant change when only the temperature is explicitly forced, as observed from the results of M5. This case produces an accurate temperature spectrum in the core of the domain and yields an improved spectrum near the wall. Additionally forcing the velocity significantly improves the velocity spectra, as is observed for cases M6 and M7. Here, we observe good agreement for the velocity and the temperature across all length scales in the center of the domain. In particular, a definite improvement is observed when comparing the temperature spectrum to those of M1-4. Near the wall, the horizontal velocity and the temperature are both captured accurately, while the vertical velocity still deviates for wavenumbers larger than 15. The similarity between the spectra obtained for M6 and M7 indicates that the heat flux correction described in Section \ref{sec:heat_transport_correction} does indeed not lead to significant changes in the spectra.

\subsection{Flow statistics}\label{sec:flow_statistics}
\begin{figure}
    \centering
    \includegraphics[width=\textwidth]{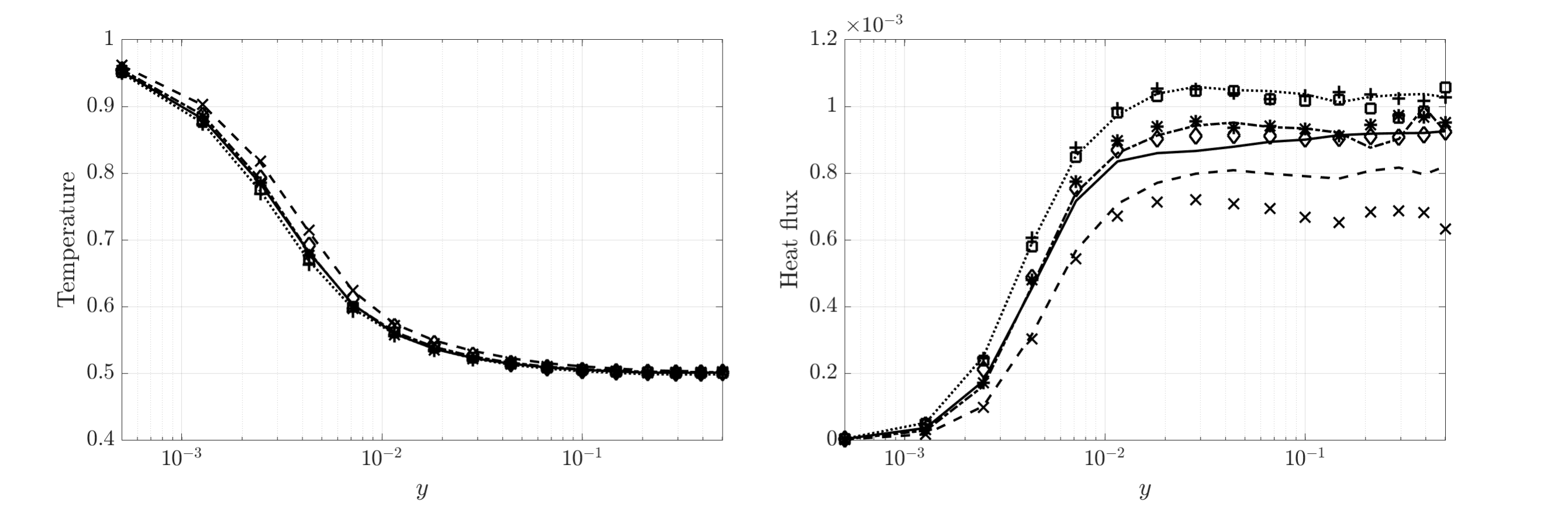}
    \caption{Comparison of the time-averaged temperature (left) and time-averaged heat flux (right) measured along horizontal cross-sections of the domain and displayed as a function of the wall-normal distance. The solid line shows the mean values of the filtered DNS, the model results are displayed using the symbols in table \ref{tab:config}.}
    \label{fig:mean_temp}
\end{figure}

The mean temperature and mean heat flux are displayed in figure \ref{fig:mean_temp} as a function of the wall-normal distance. All models except M5 efficiently mitigate the small mean temperature discrepancy between M0 and the reference. 

The mean heat flux of the no-model M0 case is consistently too low, which is a direct result of underestimating the vertical velocity. Applying the large-scale velocity forcing as done in cases M1 and M3 yields an improved heat flux. In particular, the measured heat flux near the wall shows good agreement with the reference. The mean heat flux is consistently overestimated when the velocity is forced at all wavenumbers, which is the case for M2, M4, and M6. Comparison of the results of M6 with M7 establishes that the heat flux correction described in Section \ref{sec:heat_transport_correction} ensures a better prediction of the mean heat flux. Finally, only imposing the temperature spectrum deteriorates the measured heat flux, as shown by the results of M5.

These observations in combination with the energy spectra of the previous subsection expose the simplifying model assumptions discussed in Section \ref{sec:model_description}. Despite accurate energy spectra of all variables, the M6 model does not yield an accurate heat flux. This indeed suggests that the energy spectra alone do not provide sufficiently strict modeling criteria for obtaining accurate coarse-grid numerical simulations, and instead benefit from additional cross-variable constraints such as the imposed heat flux.

\begin{figure}
    \centering
    \includegraphics[width=\textwidth]{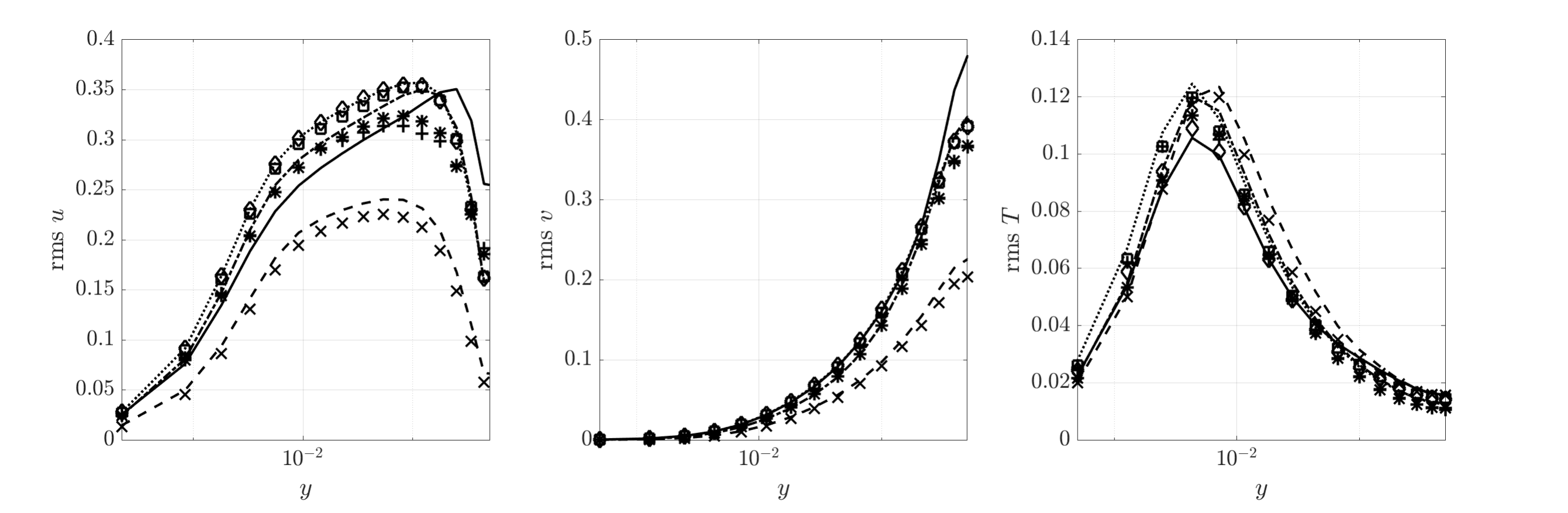}
    \caption{Root mean square (rms) of the horizontal velocity (left), vertical velocity (middle), and temperature (right), measured along horizontal cross-sections of the domain and displayed as a function of the wall-normal distance. The solid line shows the rms values of the filtered DNS, the model results are displayed using the symbols in table \ref{tab:config}.}
    \label{fig:rms}
\end{figure}

The rms of the velocity components are shown in the left and middle panels of figure \ref{fig:rms} as a function of the wall-normal distance. A strong reduction of the turbulent intensity of the velocity is observed for the no-model M0 results. Similar to previous observations for case M5, only forcing the temperature does not lead to improvements in the rms of the velocity. All other model configurations lead to a comparable improvement in the rms of the horizontal velocity. A slight difference between the stochastically forced and deterministically forced solutions may be distinguished in the rms profiles of the horizontal velocity, visible in the results of M3-4. Comparable results are observed for the rms of the vertical velocity, where all models except M0 and M5 display good agreement with the reference.

The average temperature fluctuations are shown in the right panel of figure \ref{fig:rms}. We observe that all model configurations except M0 and M5 predict the wall-normal distance of the peak of the fluctuations accurately. However, the model overestimates the maximal predicted rms by $7.5\%$ to $18\%$.

\subsection{Total kinetic energy and heat flux}\label{sec:global_stats}
\begin{figure}
    \centering
    \includegraphics[width=\textwidth]{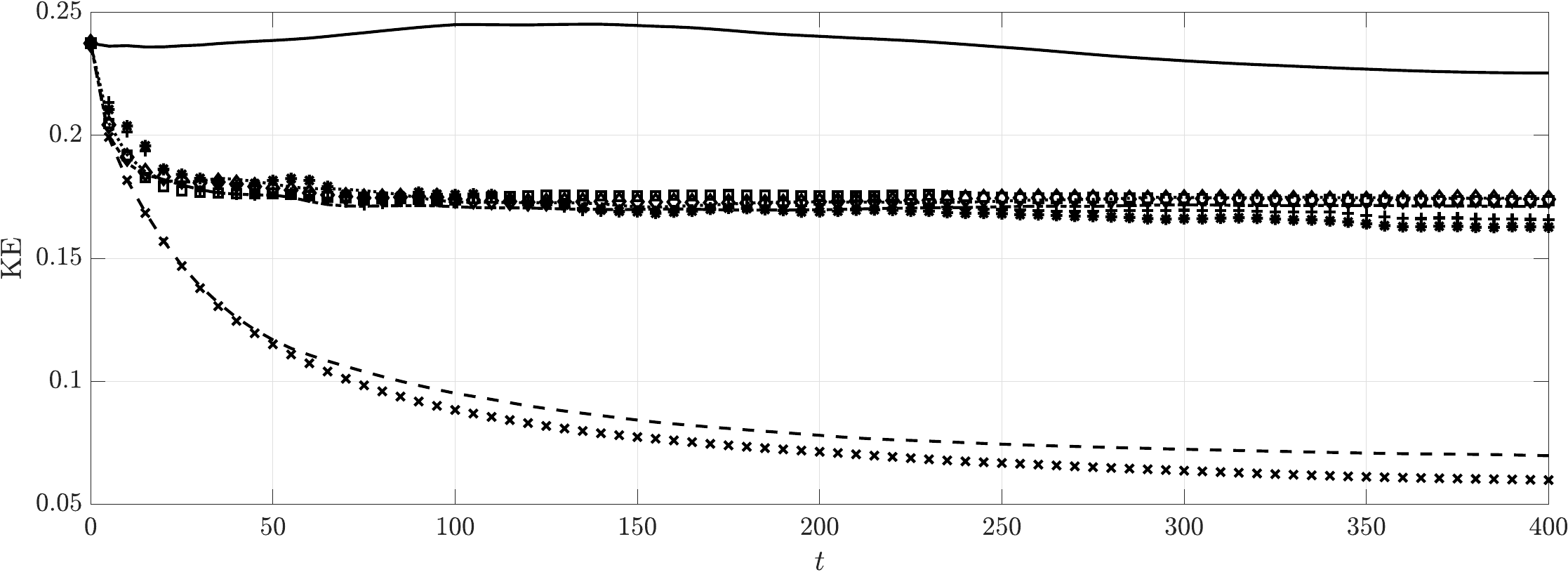}
    \caption{Comparison of the rolling mean of the kinetic energy (KE) over time. The solid line shows the KE of the filtered DNS, and the model results are displayed using the symbols in table \ref{tab:config}.}
    \label{fig:KE}
\end{figure}

A comparison of the rolling mean of the total kinetic energy (KE) is shown in figure \ref{fig:KE}. The improvement obtained by M1-4, M6, and M7 is evident. \sre{The coarse simulations are initialized from a snapshot of the filtered DNS in the reference statistically stationary state. Due to coarsening effects, this statistically stationary state cannot be sustained in coarse simulations without including a model term. Numerical dissipation and other discretization effects lead to deviations from the filtered DNS on coarse grids, despite adopting the same physical parameters as in the high-resolution simulation. The forcing model aims to steer the coarse numerical simulation towards the reference statistically stationary state by applying a correction at each time step.  Nonetheless, this procedure is approximate and does not necessarily yield the same total kinetic energy and Nusselt number as observed in the reference. Therefore, transient behavior is observed initially until the coarse simulations settle at a different statistically steady state.} At $t=400$, the mean of the KE for \sre{M0} is approximately $31\%$ of the reference KE. Only forcing the temperature, shown by M5, deteriorates the total energy and yields roughly $27\%$ of the reference value. The other models contain between $72\%$ and $77\%$ of the reference value. It is reasonable to assume that this discrepancy is predominantly caused by the model underestimating the energy in large scales in the center of the domain, as was discussed in Section \ref{sec:spectra}.

\begin{figure}
    \centering
    \includegraphics[width=\textwidth]{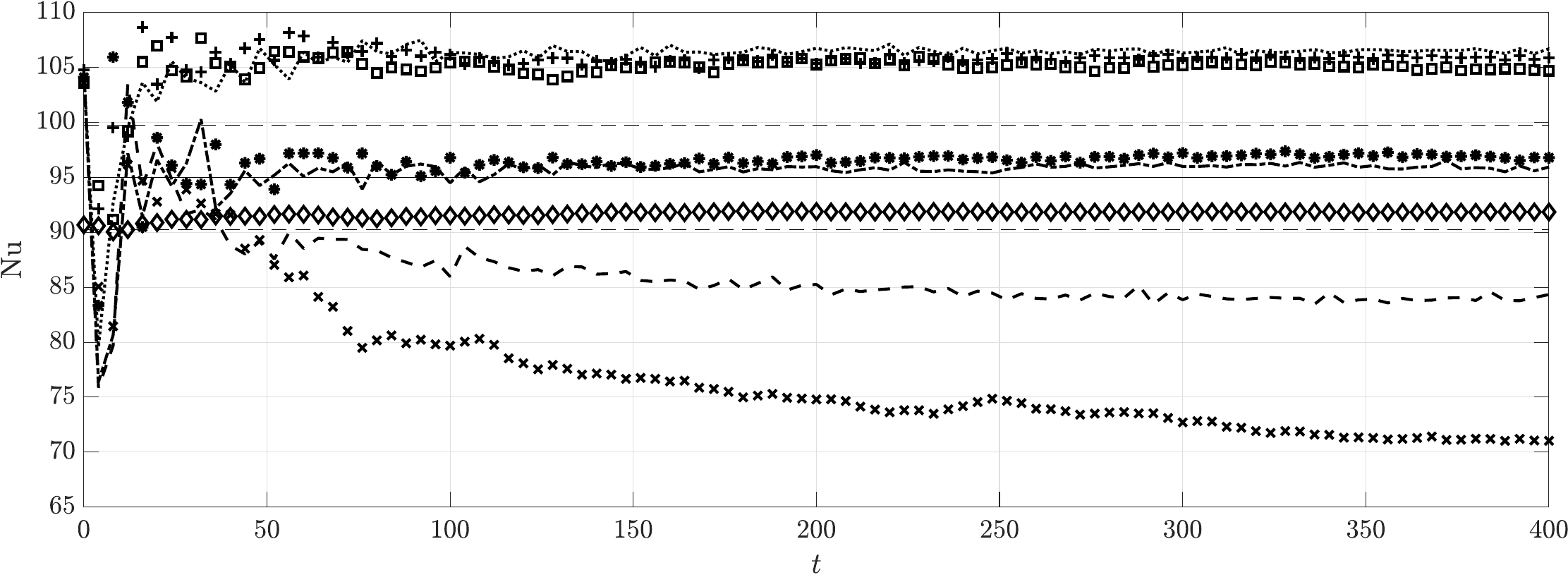}
    \caption{Comparison of the rolling mean of the Nusselt number over time. The solid line at $Nu=95$ shows the theoretically predicted value, with $5\%$ error margins given by the dashed lines. The model results are displayed using the symbols in table \ref{tab:config}.}
    \label{fig:Nusselt}
\end{figure}

A quantification of the total heat flux in the domain is provided by comparing the time-averaged Nusselt number, shown in figure \ref{fig:Nusselt}. Note that the reference value $Nu=95$ is shown with $5\%$ error margins. The no-model coarse-grid simulation leads to an underestimated heat flux resulting from the reduced velocity magnitude induced by artificial dissipation. The temperature forcing in case M5 was previously shown to not yield any improvements in the mean temperature or the velocity and does therefore not improve the Nusselt number estimate. A correction of the large-scale velocity features in configurations M1 and M3 leads to a very accurate Nusselt number estimate. Nonetheless, an accurate description of the velocity does not guarantee an accurate heat flux. This is underpinned by the results of M2, M4, and M6, which all exhibit an accurate representation of large and small velocity features, but consistently overestimate the Nusselt number.
Finally, we observe that this adverse effect is efficiently mitigated by the heat flux correction, as becomes evident from the resulting Nusselt number estimate of M7.

\begin{figure}
    \centering
    \includegraphics[width=\textwidth]{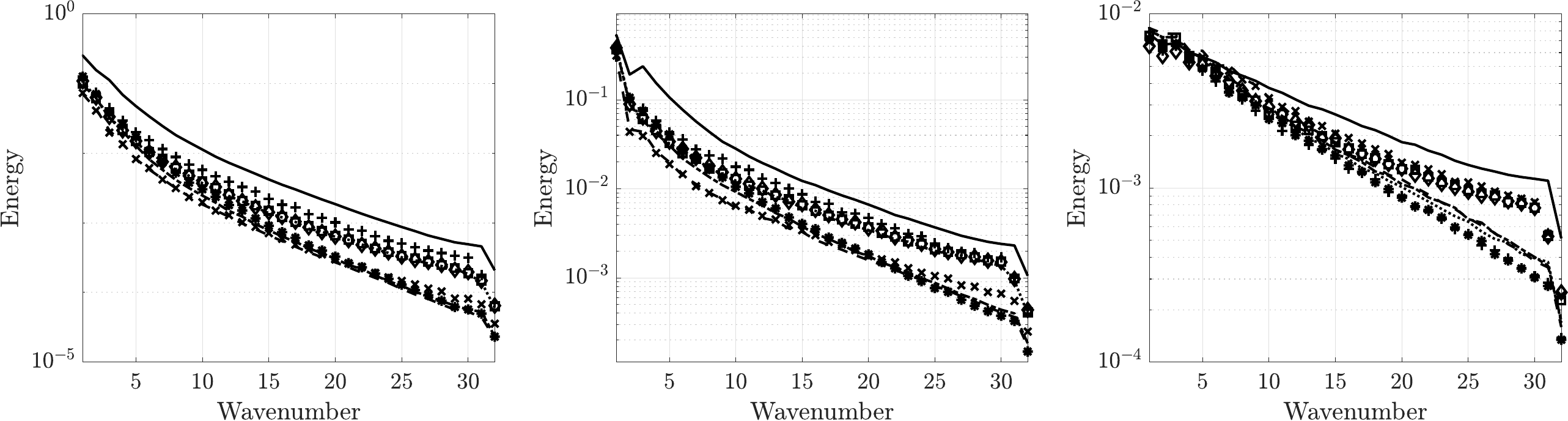}
    \caption{Time-averaged energy spectra measured along horizontal cross-sections of the domain for the horizontal velocity (left column), vertical velocity (middle column), and temperature (right column). The cross-sections are taken near the core of the domain, at $y=5.5\times10^{-1}$ for the horizontal velocity and at $y=5.0\times10^{-1}$ for the vertical velocity and the temperature. The model results are obtained by multiplying the measured correlation times by 8. The solid lines show the average spectra of the filtered DNS, the model results are displayed using the symbols in table \ref{tab:config}. }
    \label{fig:energy_spectra_tau8}
\end{figure}

\begin{figure}
\centering
\includegraphics[width=\textwidth]{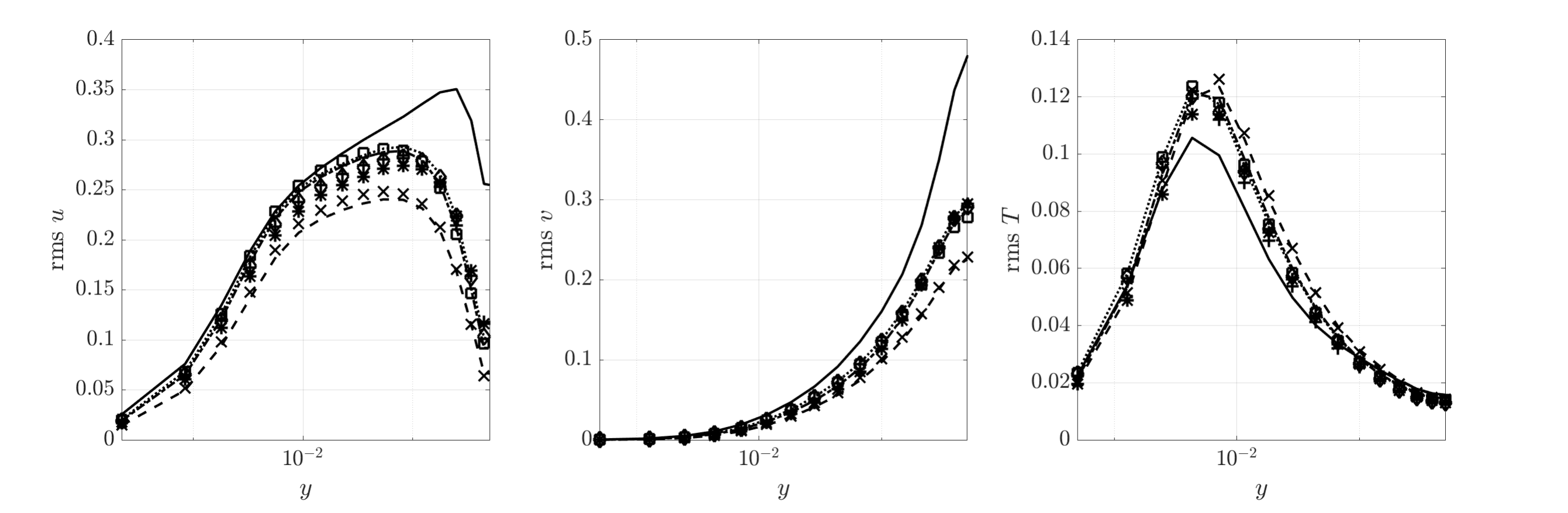}
\caption{Root mean square (rms) of the horizontal velocity (left), vertical velocity (middle), and temperature (right), measured along horizontal cross-sections of the domain and displayed as a function of the wall-normal distance. The model results are obtained by multiplying the measured correlation times by 8. The solid line shows the rms values of the filtered DNS, the model results are displayed using the symbols in table \ref{tab:config}. }
\label{fig:rms_tau8}
\end{figure}

\sre{\subsection{Robustness under variations in forcing strength}} \label{sec:tau_scaling}
\sre{The robustness of flow predictions with respect to the forcing strength is addressed next. A comparison of various forcing strengths is carried out by multiplying all measured $\tau_{k,l}$ by $0.5$, increasing the forcing strength, or by 2, 4, and 8, decreasing the forcing strength. These scalings are applied to the model configurations M0-7 described in table \ref{tab:config}. Figures \ref{fig:energy_spectra_tau8} and \ref{fig:rms_tau8} show the energy spectra in the bulk of the domain and the rms of the function of the wall-normal distance, respectively, obtained when multiplying the measured $\tau_{k,l}$ by 8. The results are only shown for this scaling since these differ the most from the results obtained with the unscaled forcing strength. \\
Increasing the forcing strength leads to numerical solutions that closely follow the reference kinetic energy spectra. As a result, the rms values as a function of the wall-normal distance also improve considerably with respect to the no-model coarse numerical simulation. However, no significant differences have been observed when compared to the results obtained with the original (unscaled) values of $\tau_{k,l}$. Decreasing the forcing strength leads to coarse numerical simulations that follow the reference energy spectra less closely, but nonetheless improve upon the no-model coarse simulation. This is displayed in Fig. \ref{fig:energy_spectra_tau8}. This result is also reflected in the rms as illustrated in Fig. \ref{fig:rms_tau8}.
}

\begin{figure}
\centering
\includegraphics[width=\textwidth]{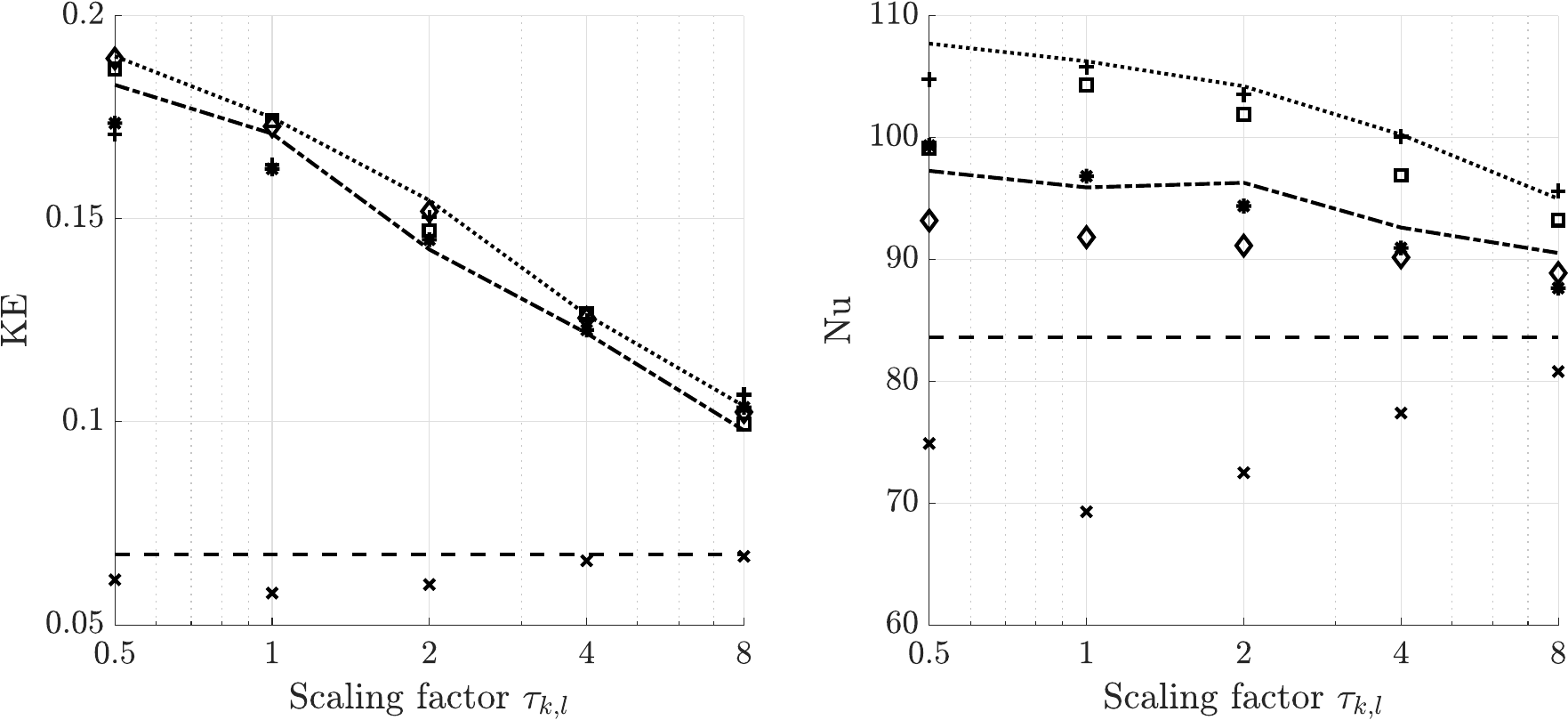}
    \caption{Comparison of the final rolling mean values of the total kinetic energy (KE, left) and Nusselt number (right) for for different scalings of the correlation times $\tau_{k,l}$ used in the model. The model results are displayed using the symbols in table \ref{tab:config}.    
    }
    \label{fig:NuKE_tau}
\end{figure}

\sre{Changing the forcing strength has a noticable effect on the total kinetic energy and the measured Nusselt number. The time-averages of the total kinetic energy and Nusselt number are depicted in Fig. \ref{fig:NuKE_tau} as a function of the scaling of the forcing strength. All model configurations except M5, where only the temperature is explicitly forced, show a decrease of the kinetic energy and of the Nusselt number as the forcing strength is decreased.  
}

\begin{figure}
\centering
\includegraphics[width=\textwidth]{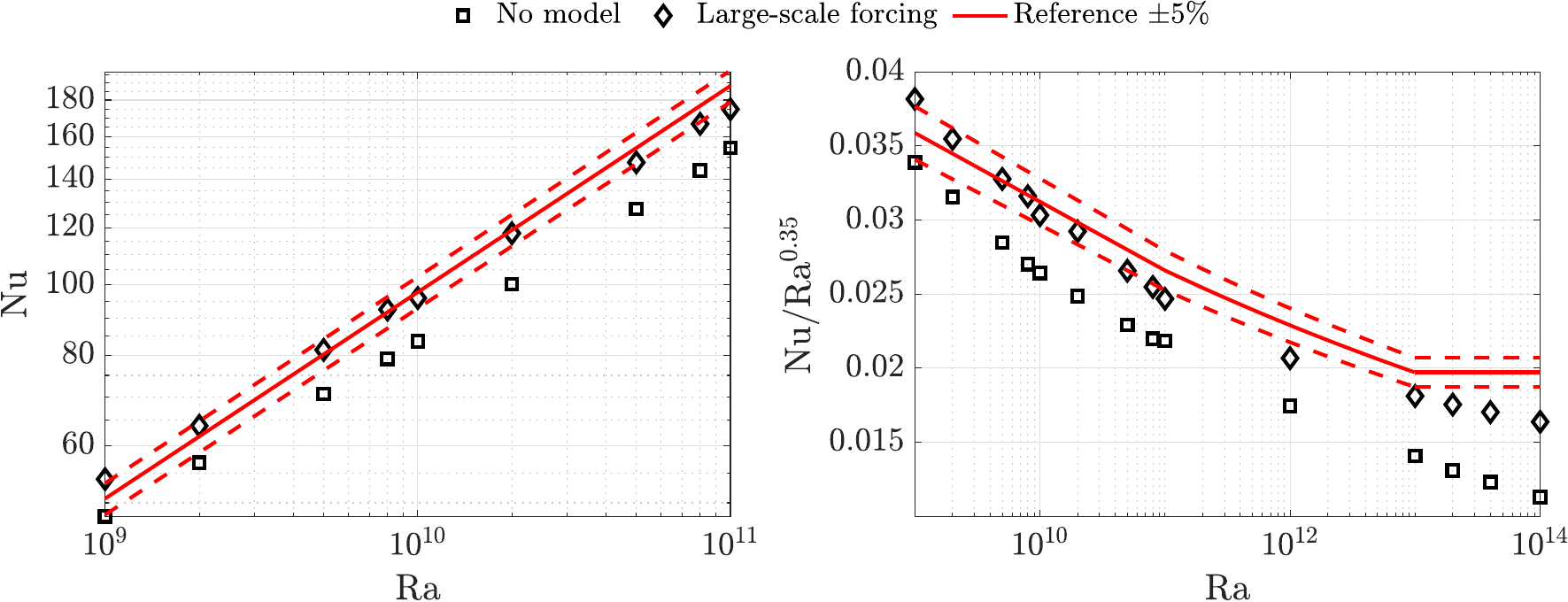}
\caption{Measured Nusselt numbers as a function of the Rayleigh number. The solid red line shows reference Nusselt number predictions from literature, with 5\% error margens given by the dashed lines. The no-model and large-scale forcing results are obtained with configurations M0 and M1, respectively. The reference scaling exponents are $0.285$ for $10^9\leq \mathrm{Ra}\leq 10^{13}$ and $0.35$ for $\mathrm{Ra}\geq 10^{13}$.}
\label{fig:Nu_Ra_scaling}
\end{figure}

\sre{\subsection{Generalization to other Rayleigh numbers}}\label{sec:Ra_scaling}
\sre{We now turn our attention to the model performance at different Rayleigh numbers.  Increasing the Rayleigh number in three-dimensional Rayleigh-B\'enard convection affects the velocity spectra and the temperature spectra along horizontal cross-sections. In \cite{de2017scalings} a range of Rayleigh numbers between $7\times 10^4$ and $2\times 10^6$ is considered at $\text{Pr}=0.71$ in a large aspect ratio box. An increase of high-wavenumber excitation is reported for kinetic energy, temperature fluctuations, heat flux, and kinetic energy dissipation, and is observed near the top and bottom plates and in the bulk of the domain. The maximum energy dissipation is found to shift towards higher wavenumbers as the Rayleigh number is increased. This might suggest that a coarse simulation at larger Rayleigh numbers benefits from increased forcing at larger wavenumbers. 

    	When increasing the Rayleigh number, the mean velocity and temperature profiles in two-dimensional Rayleigh-B\'enard convection behave similarly when normalized by the friction velocity and corresponding characteristic temperature, respectively \cite{zhu2018transition}. In the range between $\text{Ra=}10^{11}$ to $\text{Ra}=10^{14}$, the mean profiles displayed the same behavior in the viscous sublayer near the wall at each simulated Rayleigh number. This is followed by a log layer for the velocity and a flat region for the temperature. These results indicate that the mean profiles do undergo quantitative changes but no qualitative changes in this range of Rayleigh numbers. Therefore, a model calibrated at a particular Rayleigh number might still have merit when different Rayleigh numbers are considered in coarse numerical simulations. 
    	
Here, we assess the Nusselt number estimates at different Rayleigh numbers using the forcing calibrated at $\text{Ra=}10^{10}$. The Nusselt number estimates in coarse simulations without a model (M0) and with deterministic large-scale momentum forcing (M1) are compared to reference values reported in \cite{johnston2009comparison} and \cite{zhu2018transition}. The reference scaling exponents for the Nusselt number as a function of the Rayleigh number are $0.285$ for $10^9\leq \mathrm{Ra}\leq 10^{13}$ and $0.35$ for $\mathrm{Ra}\geq 10^{13}$. We consider only the model configuration M1 in this comparison, since this configuration was already found to produce an accurate Nusselt number at $\text{Ra=}10^{10}$ without using explicit knowledge of the heat flux. The results are summarized in Fig. \ref{fig:Nu_Ra_scaling}, showing the measured Nusselt number as a function of the Rayleigh number. These results show that the large-scale momentum forcing yields accurate Nusselt number estimates across several decades of Rayleigh numbers without using high-resolution data for these parameter regimes. At the transition to the so-called ultimate regime, observed at $\text{Ra}=10^{13}$ \cite{zhu2018transition}, the Nusselt number estimates of the coarse numerical simulations deteriorate somewhat. The estimates lose accuracy for much larger Rayleigh numbers compared to the reference $\text{Ra}=10^{10}$. This is likely a result of the gradually accumulating quantitative flow differences as the Rayleigh number increases. Extending the range of accurate predictions provides a challenge for future work.
}

\section{Conclusions and outlook}\label{sec:Section5}
In this paper, we have proposed a data-driven model for coarse numerical fluid simulations and assessed its performance when applied to two-dimensional Rayleigh-Bénard convection. Statistical information of Fourier coefficients of a reference direct numerical simulation was used to infer model parameters, which constituted a forcing term for reproducing the reference energy spectra. The model parameters are defined such that \sre{small scales of motion are corrected strongly}. Various model configurations were applied to gain insight into the model performance, generally leading to improved results compared to using no model.

Applying the model at all wavelengths resulted in significant improvement of the spectra both near the walls and near the center of the domain, which established that the model had its desired effect on the numerical solution. 
Additionally, the application of the model was found to yield improved estimates of flow statistics. In particular, the average turbulent fluctuations and average temperature improved significantly compared to the no-model case. The total kinetic energy was found to improve upon using the model but highlighted that \sre{modeling the effects of unresolved dynamics on the large-scale flow features as independent processes might be too restrictive.} The measured total heat flux was accurately captured for several model configurations, although accurately reconstructing energy spectra was shown not to be a sufficient criterion for this purpose. The latter problem was efficiently alleviated by including a constraint on the average heat flux in the model\sre{, leading to a computationally cheap data-driven surrogate model for the highly complex flow dynamics. Applying large-scale momentum forcing in the coarse numerical simulations yielded an accurate Nusselt number estimate without requiring explicit knowledge of the heat flux. Furthermore, accurate Nusselt number estimates were obtained across two decades of Rayleigh numbers using only the forcing calibrated at $\text{Ra}=10^{10}$. }

Future work will be dedicated to expanding the proposed model by consulting Kalman filtering theory. Specifically, the interactions between the Fourier modes can be explicitly represented by including covariance estimates in the model. This would additionally serve to verify at which frequencies the Fourier modes evolve independently, which is expected to result in a better understanding of the modeling of small-scale flow features. Alternatively, spatially coherent structures can be included in the model by applying proper orthogonal decomposition to the reference data, as demonstrated in \cite{ephrati2022stochastic}. Although no assumptions are made about the numerical method or adopted coarse resolution in the formulation of the model, further numerical experiments adopting a different resolution or discretization may be carried out to assess the robustness of the model. 
\sre{Finally, we note that the presented method might be extended to three-dimensional flows. This could be achieved by extending the method presented in this paper to also cover the reconstruction of two-dimensional energy spectra. For simple domains, covered with a structured tensor grid, this could be achieved by treating each horizontal (or constant-index) cross-section of the three-dimensional domain separately. Alternatively, one can employ proper orthogonal decomposition and recover the POD spectrum as in \cite{ephrati2022stochastic} and readily extend this from two dimensions to three dimensions. }


\section*{Acknowledgements}
The authors would like to thank Erwin Luesink and Arnout Franken, at the University of Twente, and Darryl Holm and James-Michael Leahy, at the Department of Mathematics, Imperial College London, for the inspiring discussions we could have in the context of the SPRESTO project, funded by the Dutch Science Foundation (NWO) in their TOP1 program. Computational resources were made available through the computing grant `Multiscale Modeling and Simulation' provided by NWO, the Dutch National Science Foundation. These are gratefully acknowledged. \\
This work is part of the SPRESTO project, funded by the Dutch Science Foundation (NWO) in their TOP1 program

\section*{Declaration of interests}
The authors report no conflict of interest.

\section*{Data availability statement}
The data that support the findings of this study are openly available in Zenodo at http://doi.org/10.5281/zenodo.7966196.


\bibliographystyle{plain}
\bibliography{jfm}

\begin{thebibliography}{10}

\bibitem{ahlers2009heat}
Guenter Ahlers, Siegfried Grossmann, and Detlef Lohse.
\newblock Heat transfer and large scale dynamics in turbulent
  rayleigh-b{\'e}nard convection.
\newblock {\em Reviews of modern physics}, 81(2):503, 2009.

\bibitem{altaf2015downscaling}
MU~Altaf, ES~Titi, OM~Knio, L~Zhao, MF~McCabe, and I~Hoteit.
\newblock Downscaling the 2d b{\'e}nard convection equations using continuous
  data assimilation.
\newblock {\em arXiv preprint arXiv:1512.04671}, 2015.

\bibitem{azouani2014continuous}
Abderrahim Azouani, Eric Olson, and Edriss~S Titi.
\newblock Continuous data assimilation using general interpolant observables.
\newblock {\em Journal of Nonlinear Science}, 24:277--304, 2014.

\bibitem{azouani2013feedback}
Abderrahim Azouani and Edriss~S Titi.
\newblock Feedback control of nonlinear dissipative systems by finite
  determining parameters-a reaction-diffusion paradigm.
\newblock {\em arXiv preprint arXiv:1301.6992}, 2013.

\bibitem{beck2019deep}
Andrea Beck, David Flad, and Claus-Dieter Munz.
\newblock Deep neural networks for data-driven les closure models.
\newblock {\em Journal of Computational Physics}, 398:108910, 2019.

\bibitem{blomker2013accuracy}
Dirk Bl{\"o}mker, Kody Law, Andrew~M Stuart, and Konstantinos~C Zygalakis.
\newblock Accuracy and stability of the continuous-time 3dvar filter for the
  navier--stokes equation.
\newblock {\em Nonlinearity}, 26(8):2193, 2013.

\bibitem{fedderikbos2007}
F.~van~der Bos, JJW van~der Vegt, and B.J. Geurts.
\newblock A multi-scale formulation for compressible turbulent flows suitable
  for general variational discretization techniques.
\newblock {\em Computer methods in applied mechanics and engineering},
  196(29-30):2863--2875, 2007.

\bibitem{cifani2018highly}
P~Cifani, JGM Kuerten, and BJ~Geurts.
\newblock Highly scalable {DNS} solver for turbulent bubble-laden channel flow.
\newblock {\em Computers \& Fluids}, 172:67--83, 2018.

\bibitem{daley1992estimating}
Roger Daley.
\newblock Estimating model-error covariances for application to atmospheric
  data assimilation.
\newblock {\em Monthly weather review}, 120(8):1735--1746, 1992.

\bibitem{de2017scalings}
Arnab~K De, Vinayak Eswaran, and Pankaj~K Mishra.
\newblock Scalings of heat transport and energy spectra of turbulent
  rayleigh-b{\'e}nard convection in a large-aspect-ratio box.
\newblock {\em International Journal of Heat and Fluid Flow}, 67:111--124,
  2017.

\bibitem{ephrati2023qeuler}
Sagy Ephrati, Paolo Cifani, Milo Viviani, and Bernard Geurts.
\newblock Data-driven stochastic spectral modeling for coarsening of the
  two-dimensional euler equations on the sphere.
\newblock {\em arXiv preprint arXiv:2304.12007}, 2023.

\bibitem{ephrati2023data}
Sagy~R Ephrati, Paolo Cifani, Erwin Luesink, and Bernard~J Geurts.
\newblock Data-driven stochastic lie transport modelling of the 2d euler
  equations.
\newblock {\em Journal of Advances in Modeling Earth Systems}, page
  e2022MS003268, 2023.

\bibitem{ephrati2022computational}
Sagy~R. Ephrati, Erwin Luesink, Golo Wimmer, Paolo Cifani, and Bernard~J.
  Geurts.
\newblock Computational modeling for high-fidelity coarsening of shallow water
  equations based on subgrid data.
\newblock {\em Multiscale Modeling \& Simulation}, 20(4):1468--1489, 2022.

\bibitem{ephrati2022stochastic}
SR~Ephrati, P~Cifani, and BJ~Geurts.
\newblock Stochastic data-driven pod-based modeling for high-fidelity
  coarsening of two-dimensional rayleigh-b{\'e}nard turbulence.
\newblock In {\em 13th ERCOFTAC Workshop on Direct \& Large Eddy Simulation
  2022}. ERCOFTAC, 2022.

\bibitem{farhat2015continuous}
Aseel Farhat, Michael~S Jolly, and Edriss~S Titi.
\newblock Continuous data assimilation for the 2{D} {B}{\'e}nard convection
  through velocity measurements alone.
\newblock {\em Physica D: Nonlinear Phenomena}, 303:59--66, 2015.

\bibitem{frederiksen2006dynamical}
Jorgen~S Frederiksen and Steven~M Kepert.
\newblock Dynamical subgrid-scale parameterizations from direct numerical
  simulations.
\newblock {\em Journal of the atmospheric sciences}, 63(11):3006--3019, 2006.

\bibitem{geurts2022book}
Bernard Geurts.
\newblock {\em Direct and Large-Eddy simulation}, volume
  https://doi.org/10.1515/9783110532364.
\newblock De Gruyter - Computational Science and Engineering, 2022.

\bibitem{geurtsholm2002}
Bernard~J. Geurts and Darryl Holm.
\newblock {\em Alpha-modeling strategy for LES of turbulent mixing}.
\newblock Springer - Turbulent flow computation, 2002.

\bibitem{geurts2005numerically}
Bernard~J Geurts and Fedderik van~der Bos.
\newblock Numerically induced high-pass dynamics in large-eddy simulation.
\newblock {\em Physics of fluids}, 17(12):125103, 2005.

\bibitem{geurts2003elements}
Bernardus~J Geurts.
\newblock {\em Elements of direct and large eddy simulation}.
\newblock RT Edwards, Inc, 2003.

\bibitem{geurtsholm2003}
B.J. Geurts and D.D. Holm.
\newblock Regularization modeling for large-eddy simulation.
\newblock {\em Physics of fluids}, 15(1):L13--L16, 2003.

\bibitem{ghil1991data}
Michael Ghil and Paola Malanotte-Rizzoli.
\newblock Data assimilation in meteorology and oceanography.
\newblock In {\em Advances in geophysics}, volume~33, pages 141--266. Elsevier,
  1991.

\bibitem{harlim2008filtering}
J~Harlim and AJ~Majda.
\newblock Filtering nonlinear dynamical systems with linear stochastic models.
\newblock {\em Nonlinearity}, 21(6):1281, 2008.

\bibitem{hartmann2001tropical}
Dennis~L Hartmann, Leslie~A Moy, and Qiang Fu.
\newblock Tropical convection and the energy balance at the top of the
  atmosphere.
\newblock {\em Journal of Climate}, 14(24):4495--4511, 2001.

\bibitem{heyder2021echo}
Florian Heyder and J{\"o}rg Schumacher.
\newblock Echo state network for two-dimensional turbulent moist
  rayleigh-b{\'e}nard convection.
\newblock {\em Physical Review E}, 103(5):053107, 2021.

\bibitem{higham2001algorithmic}
Desmond~J Higham.
\newblock An algorithmic introduction to numerical simulation of stochastic
  differential equations.
\newblock {\em SIAM review}, 43(3):525--546, 2001.

\bibitem{johnston2009comparison}
Hans Johnston and Charles~R Doering.
\newblock Comparison of turbulent thermal convection between conditions of
  constant temperature and constant flux.
\newblock {\em Physical review letters}, 102(6):064501, 2009.

\bibitem{kadanoff2001turbulent}
Leo~P. Kadanoff.
\newblock Turbulent heat flow: Structures and scaling.
\newblock {\em Physics Today}, 54(8):34--39, 2001.

\bibitem{kim1985application}
John Kim and Parviz Moin.
\newblock Application of a fractional-step method to incompressible
  navier-stokes equations.
\newblock {\em Journal of computational physics}, 59(2):308--323, 1985.

\bibitem{kooij2018comparison}
Gijs~L Kooij, Mikhail~A Botchev, Edo~MA Frederix, Bernard~J Geurts, Susanne
  Horn, Detlef Lohse, Erwin~P van~der Poel, Olga Shishkina, Richard~JAM
  Stevens, and Roberto Verzicco.
\newblock Comparison of computational codes for direct numerical simulations of
  turbulent rayleigh--b{\'e}nard convection.
\newblock {\em Computers \& Fluids}, 166:1--8, 2018.

\bibitem{kraichnan1962turbulent}
Robert~H Kraichnan.
\newblock Turbulent thermal convection at arbitrary prandtl number.
\newblock {\em The Physics of Fluids}, 5(11):1374--1389, 1962.

\bibitem{kunnen2009}
R.P.J. Kunnen, B.J. Geurts, and H.J.H. Clercx.
\newblock Turbulence statistics and energy budget in rotating
  rayleigh–bénard convection.
\newblock {\em European Journal of Mechanics-B/Fluids}, 28(4):578--589, 2009.

\bibitem{kurz2023deep}
Marius Kurz, Philipp Offenh{\"a}user, and Andrea Beck.
\newblock Deep reinforcement learning for turbulence modeling in large eddy
  simulations.
\newblock {\em International Journal of Heat and Fluid Flow}, 99:109094, 2023.

\bibitem{langford1999optimal}
Jacob~A Langford and Robert~D Moser.
\newblock Optimal {LES} formulations for isotropic turbulence.
\newblock {\em Journal of fluid mechanics}, 398:321--346, 1999.

\bibitem{leonard1979stable}
Brian~P Leonard.
\newblock A stable and accurate convective modelling procedure based on
  quadratic upstream interpolation.
\newblock {\em Computer methods in applied mechanics and engineering},
  19(1):59--98, 1979.

\bibitem{lorenc2005does}
Andrew~C Lorenc and F~Rawlins.
\newblock Why does 4d-var beat 3d-var?
\newblock {\em Quarterly Journal of the Royal Meteorological Society: A journal
  of the atmospheric sciences, applied meteorology and physical oceanography},
  131(613):3247--3257, 2005.

\bibitem{majda2012filtering}
Andrew~J Majda and John Harlim.
\newblock {\em Filtering complex turbulent systems}.
\newblock Cambridge University Press, 2012.

\bibitem{marshall1999open}
John Marshall and Friedrich Schott.
\newblock Open-ocean convection: Observations, theory, and models.
\newblock {\em Reviews of geophysics}, 37(1):1--64, 1999.

\bibitem{maulik2019subgrid}
Romit Maulik, Omer San, Adil Rasheed, and Prakash Vedula.
\newblock Subgrid modelling for two-dimensional turbulence using neural
  networks.
\newblock {\em Journal of Fluid Mechanics}, 858:122--144, 2019.

\bibitem{pandey2022direct}
Sandeep Pandey, Philipp Teutsch, Patrick M{\"a}der, and J{\"o}rg Schumacher.
\newblock Direct data-driven forecast of local turbulent heat flux in
  rayleigh--b{\'e}nard convection.
\newblock {\em Physics of Fluids}, 34(4), 2022.

\bibitem{piomelli2015}
U.~Piomelli, A.~Rouhi, and Bernard~J. Geurts.
\newblock A grid-independent length scale for large-eddy simulations.
\newblock {\em Journal of fluid mechanics}, 766:499--527, 2015.

\bibitem{pope2000turbulent}
Stephen~B Pope and Stephen~B Pope.
\newblock {\em Turbulent flows}.
\newblock Cambridge university press, 2000.

\bibitem{rai1991direct}
Man~Mohan Rai and Parviz Moin.
\newblock Direct simulations of turbulent flow using finite-difference schemes.
\newblock {\em Journal of computational physics}, 96(1):15--53, 1991.

\bibitem{rouhi2016}
A.~Rouhi, U.~Piomelli, and B.J. Geurts.
\newblock Dynamic subfilter-scale stress model for large-eddy simulations.
\newblock {\em Physical review fluids}, 1(4):044401, 2016.

\bibitem{sagaut2006large}
Pierre Sagaut.
\newblock {\em Large eddy simulation for incompressible flows: an
  introduction}.
\newblock Springer Science \& Business Media, 2006.

\bibitem{stevens2018turbulent}
Richard~JAM Stevens, Alexander Blass, Xiaojue Zhu, Roberto Verzicco, and Detlef
  Lohse.
\newblock Turbulent thermal superstructures in rayleigh-b{\'e}nard convection.
\newblock {\em Physical review fluids}, 3(4):041501, 2018.

\bibitem{van2015pencil}
Erwin~P Van Der~Poel, Rodolfo Ostilla-M{\'o}nico, John Donners, and Roberto
  Verzicco.
\newblock A pencil distributed finite difference code for strongly turbulent
  wall-bounded flows.
\newblock {\em Computers \& Fluids}, 116:10--16, 2015.

\bibitem{van2013comparison}
Erwin~P. van~der Poel, Richard J. A.~M. Stevens, and Detlef Lohse.
\newblock Comparison between two- and three-dimensional {R}ayleigh–{B}énard
  convection.
\newblock {\em Journal of Fluid Mechanics}, 736:177–194, 2013.

\bibitem{vlachas2020backpropagation}
Pantelis~R Vlachas, Jaideep Pathak, Brian~R Hunt, Themistoklis~P Sapsis,
  Michelle Girvan, Edward Ott, and Petros Koumoutsakos.
\newblock Backpropagation algorithms and reservoir computing in recurrent
  neural networks for the forecasting of complex spatiotemporal dynamics.
\newblock {\em Neural Networks}, 126:191--217, 2020.

\bibitem{vreman2014projection}
AW~Vreman.
\newblock The projection method for the incompressible navier--stokes
  equations: the pressure near a no-slip wall.
\newblock {\em Journal of Computational Physics}, 263:353--374, 2014.

\bibitem{vreman1997large}
Bert Vreman, Bernard Geurts, and Hans Kuerten.
\newblock Large-eddy simulation of the turbulent mixing layer.
\newblock {\em Journal of fluid mechanics}, 339:357--390, 1997.

\bibitem{zhu2018transition}
Xiaojue Zhu, Varghese Mathai, Richard~JAM Stevens, Roberto Verzicco, and Detlef
  Lohse.
\newblock Transition to the ultimate regime in two-dimensional
  {R}ayleigh-{B}{\'e}nard convection.
\newblock {\em Physical review letters}, 120(14):144502, 2018.

\end{thebibliography}

\end{document}